\documentclass[onecolumn,authoryear]{els-mrw} 

\usepackage{amsmath,amssymb,amsfonts,amsthm,makeidx,graphicx}
\usepackage{txfonts}
\usepackage{helvet}

\newcommand{\lya}{Ly$\alpha$}
\newcommand{\lyb}{Ly$\beta$}

\newcommand{\hi}{H~{\sc i}}
\newcommand{\hii}{H~{\sc ii}}
\newcommand{\heii}{He~{\sc ii}}
\newcommand{\ovi}{O~{\sc vi}}
\newcommand{\ovii}{O~{\sc vii}}
\newcommand{\nv}{N~{\sc v}}
\newcommand{\nhi}{$N_{\rm HI}$}
\newcommand{\lcdm}{$\Lambda$CDM}
\newcommand{\cm}{cm$^{-2}$}
\newcommand{\cc}{cm$^{-3}$}  
\newcommand{\kms}{km\,s$^{-1}$}

\newcommand{\dndztt}{${\rm d}\mathcal{N_{\rm abs}}/{\rm d}z$}

\newcommand{\chclusters}{``Galaxy clusters and the local Universe''} 
\newcommand{\chqsos}{``AGN across cosmic time'' }  
\newcommand{\chgasingal}{``Gas in galaxies''}  
\newcommand{\chism}{``The interstellar medium''}  
\newcommand{\chcgm}{``Circumgalactic medium''}  
\newcommand{\chgalwinds}{``Galactic winds and outflows''}  
\newcommand{\cheor}{``Reionization and its sources''}  
\newcommand{\chfirstgals}{``First galaxies''}  
\newcommand{\chgalevol}{``Cosmological simulations of galaxies''}  
\newcommand{\chcosmicweb}{``Large-scale structure and the cosmic web''}  
\newcommand{\chfrbs}{``Fast radio bursts''}  
\newcommand{\chcosmofrb}{``Cosmology with fast radio bursts''}  
\newcommand{\chgrbs}{``Gamma-ray bursts''}  
\newcommand{\chexpand}{``The expanding Universe''}  
\newcommand{\chbbn}{``Big Bang Nucleosynthesis''}  
\newcommand{\chcmb}{``The cosmic microwave background''}  
\newcommand{\chlinear}{``Linear structure formation''} 
\newcommand{\chpowerspectrum}{``Power spectra and correlation functions''}  
\newcommand{\chnonlinear}{``Non-linear structure formation''} 
\newcommand{\chbao}{``Baryon acoustic oscillations''}  
\newcommand{\chcosmolya}{``Cosmology with the Lyman-$\alpha$ forest''}  

\newcommand{\sk}{\smallskip}

\newcommand{\cloudy}{{\sc cloudy}}

\newcommand{\linetools}{{\sc linetools}}

\newcommand{\araa}{Annual Review of Astronomy and Astrophysics}
\newcommand{\apj}{The Astrophysical Journal}
\newcommand{\apjs}{The Astrophysical Journal Supplement Series}
\newcommand{\aj}{The Astronomical Journal}
\newcommand{\apjl}{The Astrophysical Journal Letters}
\newcommand{\mnras}{Monthly Notices of the Royal Astronomical Society}

\newcommand{\nat}{Nature}
\newcommand{\jcap}{Journal of Cosmology and Astroparticle Physics}
\newcommand{\prd}{Physical Review D}
\newcommand{\aap}{Astronomy and Astrophysics}
\newcommand{\aaps}{Astronomy and Astrophysics Supplement Series}
\newcommand{\aapr}{The Astronomy and Astrophysics Review}
\newcommand{\pasp}{Publications of the Astronomical Society of the Pacific}

\begin{document}

\chapter{The Intergalactic Medium}\label{chap1}

\author[1]{Nicolas Tejos}%

\address[1]{\orgname{Instituto de F\'isica, Pontificia Universidad Cat\'olica de
Valpara\'iso}, 
\orgaddress{Casilla 4059, Valpara\'iso, Chile}}

\articletag{This is a pre-print of a chapter for the Encyclopedia of Astrophysics (edited by I. Mandel, section editor S. McGee) to be published by Elsevier as a Reference Module.\\
Date: \today}

\maketitle






\begin{glossary}[Nomenclature]
\begin{tabular}{@{}lp{34pc}@{}}
AGN & Active galactic nuclei \\
BLA & Broad Lyman-$\alpha$ absorber\\
CGM & Circumgalactic medium \\
CIE & Collisional ionization equilibrium\\
CMB & Cosmic microwave background\\
COG & Curve of growth \\
DLA & Damped Lyman-$\alpha$ system\\
EoR & Epoch of Reionization \\
FRB & Fast radio burst\\
FWHM & Full width at half maximum\\
GRB & Gamma-ray burst\\
ICM & Intracluster medium \\
IGM & Intergalactic medium \\
IR & Infrared\\
ISM & Interstellar medium \\
\lcdm & $\Lambda$~cold dark matter\\
LLS & Lyman limit system \\
PIE & Photoionization equilibrium \\
SZ & Sunyaev-Zeldovich [effect]\\
UV & Ultraviolet\\
UVB & Ultraviolet background \\
WCGM & Warm circumgalactic medium\\
WHIM & Warm-hot intergalactic medium \\

\end{tabular}
\end{glossary}

\begin{abstract}[Abstract]
The intergalactic medium (IGM) comprises all the matter that lies between galaxies. Hosting the vast majority ($\gtrsim 90\%$) of the baryons in the Universe, the IGM is a critical reservoir and probe for cosmology and astrophysics, providing insights into large-scale structure formation and galaxy evolution. In this Chapter, we present an overview of the general properties of the IGM, focusing on their dependence on cosmic environment and cosmic time. Emphasis is given to the basic physical principles that allow us to model the density, temperature, and ionization state of the IGM, supported by results from cosmological hydrodynamical simulations. We also cover the foundational principles of quasar spectroscopy used to probe the IGM in absorption, with a particular focus on \hi~absorption lines. Finally, we briefly discuss future prospects and complementary observational techniques to enhance our understanding of the IGM.
\end{abstract}

\begin{keywords}
Intergalactic medium, 
galaxy evolution, 
large-scale structure of the Universe, 
quasar absorption line spectroscopy, \lya\ forest
\end{keywords}

\paragraph{Learning objectives}\sk
This Chapter will help the readers to answer the following questions:\sk
\begin{itemize}
    \item What is the intergalactic medium (IGM) and what is its importance in astrophysics and cosmology?
    \item What is the composition and ionization state of the IGM?
    \item What are the different gas phases of the IGM?
    \item How do we model the ionization and thermal state of the IGM?
    \item How do the properties of the IGM change with cosmic environment and cosmic time?
    \item How do we observe the IGM using quasar spectroscopy?
    \item How do we model absorption lines in order to infer physical properties of the IGM?
    \item What are the different types of \hi~absorption line systems depending on their column density?
    \item How can we use these observational results to constrain physical parameters of the IGM?
\item Are there other observational probes of the IGM besides quasar spectroscopy?
\end{itemize}


\section{Introduction}\label{sec:intro}

\subsection{Definition and general properties}\label{sec:definition}

The intergalactic medium (IGM) is defined as the baryonic matter and space that exists {\it between} galaxies; in this manner, the IGM definition is linked to that of galaxies. More fundamentally, the IGM and galaxies are indeed physically interlinked, such that they are affected by each other in their co-evolution across cosmic time and cosmic environments. Although the properties and nature of galaxies have been addressed extensively in the rest of this Encyclopedia, here we must emphasize that a well defined boundary between galaxy halos and the IGM does not exist. Therefore, we should be always cautious in how we establish the distinction between galaxies and the IGM, keeping in mind that this distinction may well depend on the particular study at hand. \smallskip

Here we will adopt the terminology that the IGM corresponds to the diffuse medium outside dark matter halos of galaxies, as informed by current cosmological hydrodynamical simulations (see Section~\ref{sec:hydro}). In this manner, the interface between the IGM and galaxies is considered a different medium: the so-called circumgalactic medium (CGM). The CGM is sufficiently different and complex that has gained renewed attention, specially in the context of galaxy evolution (\citealt{tumlinson2017, faucher2023}; see also Chapter~\chcgm).\smallskip

Far away from galaxy halos, the typical densities of the IGM are comparable to the mean density of the Universe. On average, the IGM has a density of just one-fourth of an hydrogen atom per cubic meter (Section~\ref{sec:densities}), far lower than the densities typically found within galaxies---for comparison, the typical density of the interestellar medium (ISM) is a few hydrogen atoms per cubic centimeter (\citealt{draine2011}; see Chapter~\chism), i.e. a factor of $10^6$ larger. In the IGM, interactions between particles are usually infrequent, and the mean free path of photons is large. As a result, the IGM can be considered nearly transparent (i.e. optically thin) to most types of radiation, and its properties are primarily governed by larger-scale cosmic phenomena rather than local interactions (see Section~\ref{sec:igmphysics}).\sk

Despite its diffuse nature, the IGM plays an important role in cosmology and galaxy evolution. According to the $\Lambda$ Cold Dark Matter (\lcdm) paradigm, baryonic matter accounts for $\sim 5\%$ of the total matter-energy content of the Universe, with dark matter comprising about $27\%$, and dark energy making up the remaining $68\%$ (\citealt{planck2020}; see Chapters~\chexpand~and~\chcmb). Out of this $\sim 5\%$ of baryons, the vast majority ($\gtrsim 90\%$) of them are located in the IGM, whereas $\lesssim 10\%$ are locked up in galaxies or galaxy clusters (\citealt{fukugita1998, shull2012}; see also Section~\ref{sec:qal}). This highlights the fundamental importance of the IGM for studying cosmic structure and baryonic and non-baryonic matter evolution (\citealt{hernquist1996, croft1998,viel2004}; see also Chapter~\chcosmolya). In the early Universe, galaxies formed from the cooling of gas in the IGM that accumulated in dark matter halos (\citealt{peebles1974, press1974, white1978, mo2010}; see also Chapter~\chfirstgals). Since then, the IGM has served as the primary reservoir of gas that continues to feed galaxies, providing the essential material for ongoing star formation, which in turn drives galaxy growth \citep{tumlinson2017, peroux2020}. This process is particularly important in the context of galaxy evolution, as the amount of available cold gas in the IGM and CGM directly influences for how long galaxies can sustain star formation (\citealt{keres2005, dekel2006, faucher2023}; see also Chapter~\chgalevol).\sk

\begin{figure}[t]
\centering
\includegraphics[width=0.9\textwidth]{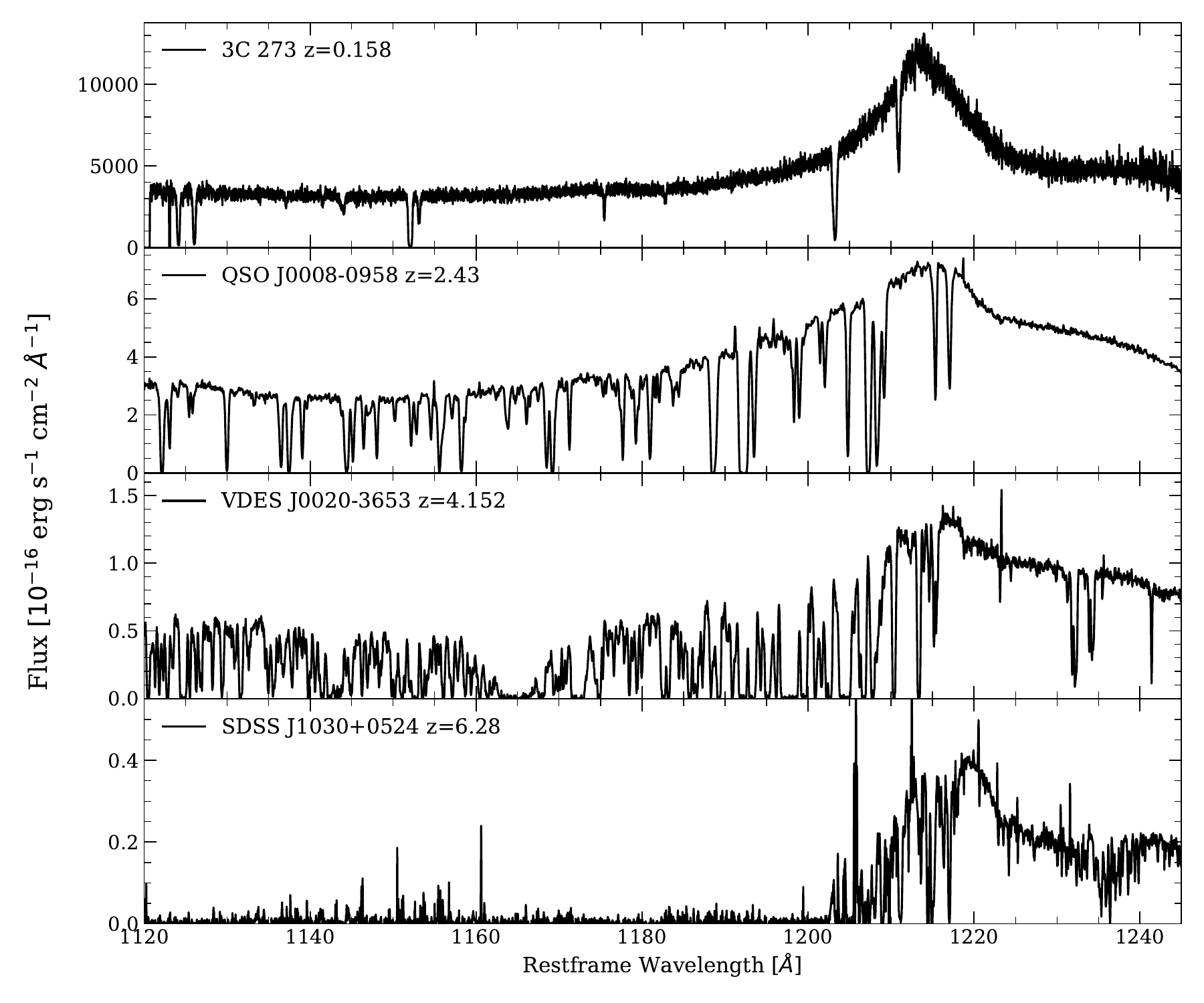}
\caption{A sample of quasar spectra covering a wide range redshifts from $0.1 \lesssim z_{\rm qso} \lesssim 6.3$ (see labels). The wavelength axis is shown in the quasars' rest-frame, such that their \lya\ emission lines are located at $\lambda^{\rm qso}_{\rm Lya}=1215.67$\,AA\AA. The \lya-forest towards these quasars is seen as series of absorption lines at wavelengths $\lambda <\lambda_{\rm Lya}^{\rm qso}$. In contrast, at $\lambda >\lambda_{\rm Lya}^{\rm qso}$ the quasar's continuum is mostly unabsorbed (although some metal ion lines are present). This figure illustrates how the \lya-forest gets more prominent with redshift, as higher redshift quasars allows us to probe it. Figure credit: K. Mart\'inez-Acosta.}
\label{fig:lyaforest}
\end{figure}

\subsection{Early discoveries and key observations}\label{sec:history}

The first direct evidence of intervening diffuse matter outside galaxies came right after the discovery of quasars (see Chapter~\chqsos) in the 1960's \citep{schmidt1963}. Optical spectroscopy of these bright cosmological sources revealed the presence of intervening metals and neutral hydrogen towards their sightlines, producing absorption lines at redshifts smaller than that of the quasar itself, $z_{\rm abs}<z_{\rm qso}$ \citep{schmidt1965, burbidge1966, lynds1971}. One of the most notable features of these $z\gtrsim 2$ quasar observations is the so-called `Lyman-$\alpha$ (\lya) forest'---a series of packed absorption lines that appear blueward of the quasar's rest-frame \lya\ emission wavelength ($\lambda_{\rm Lya}=1215.67$\,\AA) resembling `trees' in a dense forest (see Figure~\ref{fig:lyaforest}; see also \citealt{rauch1998}). It was soon recognized that such a signature corresponded to the absorption lines imprinted by a medium containing some amount of neutral hydrogen atoms (\hi), located between the quasars and the observer \citep{bahcall1965,gunn1965}.\smallskip 

Based on the inferred optical depths of these \hi\ absorption lines, \citet{gunn1965} came to an important conclusion: either the space between galaxies was nearly empty or it was highly ionized.\footnote{Otherwise the expected signal of neutral intervening medium would be that of a quasar continuum completely absorbed blueward to the rest-frame \lya\ emission (see Section~\ref{sec:qal}).} Today we know that the correct interpretation is the latter, and these kind of observations allow us to infer that only $1$ in every $\sim 10^6$ hydrogen atoms is neutral (see Section~\ref{sec:qal}); the IGM is indeed a highly ionized medium (at least at $z\lesssim 6$, i.e. after reionization; see below).\sk

Given its diffuse nature, the IGM does not emit enough photons to allow its detection in emission with current technology (as we usually do with galaxies or other luminous objects). Thus, quasar spectroscopy has remained being the prime technique for probing the IGM {\it in absorption}. By properly modeling the intervening absorption lines of neutral hydrogen\footnote{Note that \hii\ ions (free protons) are completely invisible to this technique.}
and other ions (e.g. metals), we can infer the physical properties of the IGM as a function of cosmic time and environment (see Section~\ref{sec:qal}). Ultraviolet (UV), optical, and infrared (IR) spectroscopy have been extensively used to probe \hi\ in the late ($z\lesssim 2$), intermediate ($2 \lesssim z \lesssim 5$), and early ($z \gtrsim 5$) Universe, respectively. Over the last decades, these observations, in combination with cosmological hydrodynamical simulations, have brought a significant advancement of our understanding of the IGM, providing us with a coherent picture of its evolution over cosmic time \citep{mcquinn2016}.\sk

The main goal of this Chapter is to serve as a first entry point for those interested  in studies of the IGM. We provide a general description of the physics of the IGM (Section~\ref{sec:igmphysics}) and an introduction to how we currently observe the IGM with quasar spectroscopy (Section~\ref{sec:obs}). Emphasis is given to the modeling (Section~\ref{sec:qal_model}) and interpretation (Section~\ref{sec:qal}) of intervening \hi\ absorption lines, whereas other complementary techniques and future prospects are briefly mentioned (Section~\ref{sec:final}).

\begin{figure}[t]
\centering
\includegraphics[width=.65\textwidth, angle=90]{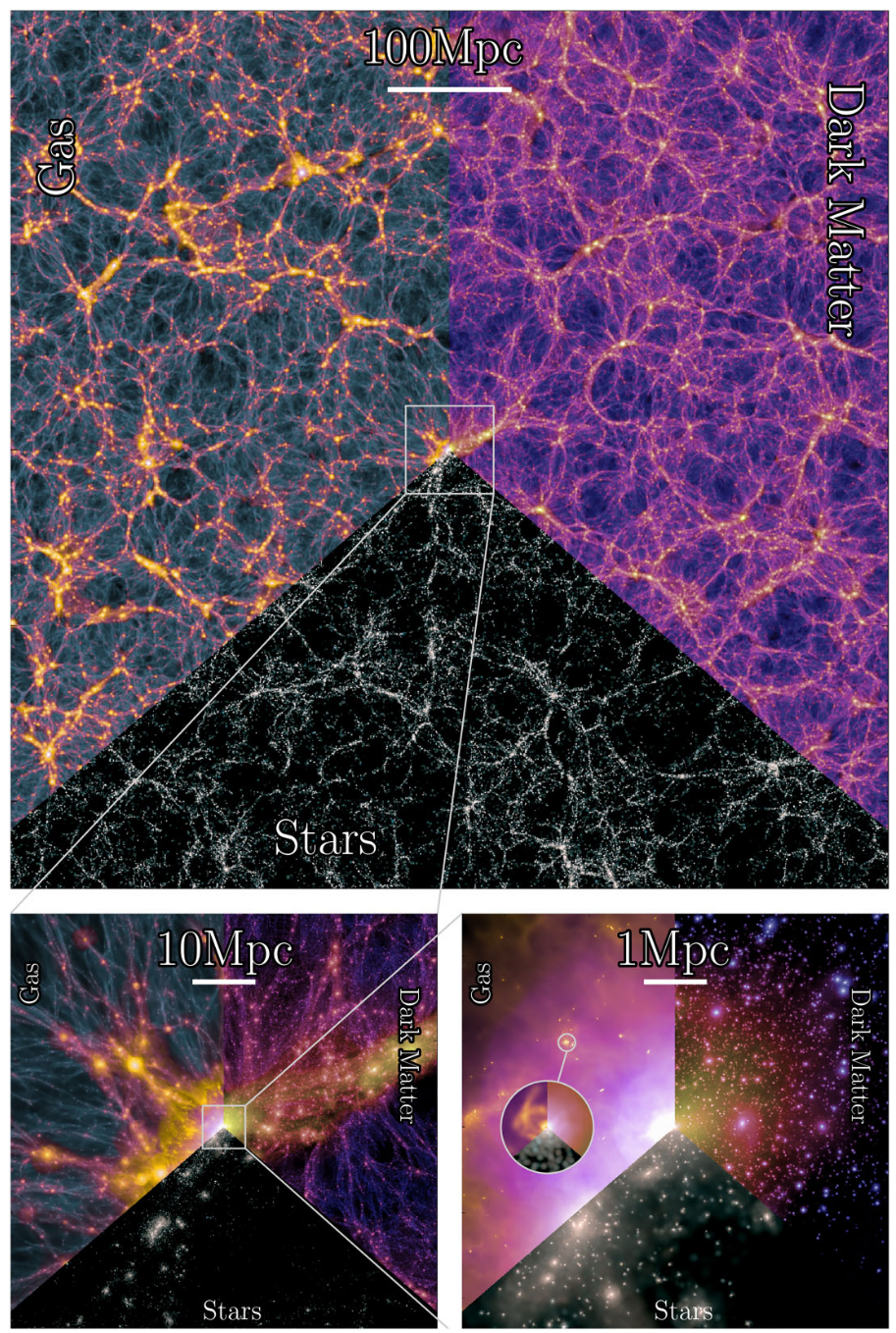}
\caption{A visualization of the output of the state-of-the-art cosmological hydrodynamical MillenniumTNG at $z=0$ \citep{pakmor2023}. In each sub-panel slices of $10$\,Mpc in thickness are shown for different components: dark matter (top left), diffuse baryons (gas; bottom left), and stars (right). The left sub-panel covers $740$\,Mpc on a side, while the insets are a factor of 10 (bottom right) and 100 (top right) smaller. A zoomed in view centred in single galaxy halo (few $100$\,kpc in diameter) is shown (circles in the top right panel). In the vast volumes of the Universe, diffuse baryons dominate and galaxies (including their stars, planets, dust, etc.) are just extremely small and dense pockets of baryons at the centers of dark matter halos. Yet, galaxies can still influence their surroundings (e.g. the CGM and IGM) via astrophysical processes occurring within them. Figure from \citet{pakmor2023}.
}
\label{fig:milleniumTNG}
\end{figure}

\section{The physics of the IGM}
\label{sec:igmphysics}

\subsection{Modeling baryons in a cosmic web shaped by dark matter}

Dark matter is the dominant matter component of the Universe that provides the structural skeleton on the largest scales \citep{white1978, mo2010}. On cosmic (Mpc) scales, dark matter forms a `cosmic web' that extend across the Universe (\citealt{bond1996}; see also Chapter~\chcosmicweb). Baryonic matter is gravitationally attracted to these dark matter structures, accumulating in denser regions, including sheets, filaments and nodes (see Figure~\ref{fig:milleniumTNG} and Section~\ref{sec:hydro}). \sk

At smaller, galaxy halo scales (100~kpc), dark matter's role becomes particularly important for the formation of stars and galaxies by drawing baryons from the IGM into their very centers. In galaxy formation models, dark matter halos act as sites where baryonic gas can cool and condense to form stars (\citealt{press1974, mo2010}; see also Chapter~\chgalevol). Along their evolution, galaxies can expel or recycle their baryons via galactic `feedback' processes---e.g. stellar winds, and active galactic nuclei (AGN)---, which heat and enrich their own CGM, and that can also reach the IGM (\citealt{veilleux2020}; see also Chapter~\chgalwinds; ). Moreover, high energy photons from hot stars and/or quasars, eventually escape from galaxies injecting further energy into the CGM and IGM (\citealt{robertson2022}; see also see Chapter~\cheor). These processes change the temperature, ionization state, and chemical composition of the CGM and the IGM, which in turn also influences and affects the future evolution of galaxies \citep{somerville2015,mcquinn2016, tumlinson2017, faucher2023}. \smallskip

Despite these complex processes occurring within and near galaxies, the IGM remains relatively simple to model. The key physical ingredients are density, temperature, and ionization state, although  metallicity can also be important in some contexts (not extensively discussed here). Therefore, by determining densities as the Universe evolve, and by identifying the main sources of ionization, heating, and cooling, we can, in principle, track the evolution of the physical state of the IGM \citep{meiksin2009, mcquinn2016}.\sk

\subsection{The Epoch of Reionization}\sk 

The Epoch of Reionization (EoR) is the period in cosmic time where the Universe transitioned from being mostly neutral to being mostly ionized. It is thought that this occurred because high-energy photons produced within the first galaxies (e.g. from young hot stars and/or AGNs) could eventually escape them and started to ionize the hydrogen in the IGM (\citealt{robertson2022}; see also Chapter~\cheor). Although the exact mechanism(s) of this process is still unknown, observations and current modeling indicate that the EoR lasted between $6\lesssim z \lesssim 15$; the several transmission spikes observed in the highly absorbed \lya-forest of $z\sim 6$ quasars are a clear sign that the EoR has already ended at these redshifts (\citealt{fan2006}; see also Section~\ref{sec:gunn-peterson}). Therefore, the ionization state of the IGM is a function of redshift: before EoR ($z\gtrsim 15$) it was mostly neutral, whereas after EoR ($z\lesssim 6$) it became and stayed a highly ionized medium (\citealt{mcquinn2016}; see also Section~\ref{sec:ionization}).\sk

\subsection{Densities and cosmic environments}\sk
\label{sec:densities}

The IGM consists of $75\%$ hydrogen and $25\%$ helium, with just traces of heavier elements (\citealt{cyburt2016}; see also Chapter~\chbbn). Initially, the matter distribution was roughly uniform (see Chapter~\chlinear), but as the Universe evolved, the cosmic web structure became more and more complex (see Chapter~\chnonlinear). For tracking densities and environments, it is convenient to define the density contrast, $\delta$, as:
\begin{align}\label{eq:overdensity}
    \delta \equiv \frac{\rho - \langle \rho \rangle}{\langle \rho \rangle} \ \rm{,}
\end{align}

\noindent where $\rho$ is the actual density at a given location and $\langle \rho \rangle$ is the average density of the Universe. Because the Universe is expanding, $\delta$, $\rho$ and $\langle \rho \rangle$ are function of cosmic time, or redshift, $z$. In a flat \lcdm\ Universe, $\langle \rho \rangle$ is equal to the critical density, $\rho_{\rm c}$, which at $z=0$, corresponds to $\rho_{{\rm c, 0}} \approx 8.6 \times 10^{-30}$\,g\,cm$^{-3}$ \citep{planck2020}.\footnote{The critical density of the Universe at $z=0$ is defined as $\rho_{\rm c,0} \equiv \frac{3H_0^2}{8\pi G}$, where $H_0$ is the Hubble constant, and $G$ is the gravitational constant.} Therefore, the cosmic baryon density at $z=0$ corresponds to:
\begin{align}\label{eq:overdensity2}
    \rho_{\rm b,0} = \Omega_{\rm b,0} \ \rho_{\rm c,0} \approx 4.2 \ \times 10^{-31} \ \rm{g\,cm}^{-3} \ \rm{,}
\end{align}

\noindent where $\Omega_{\rm b,0}\approx 0.05$ is the cosmological baryon density parameter at $z=0$ (\citealt{planck2020,fields2020}; see also Chapter~\chbbn). In terms of number density of hydrogen atoms (the dominant baryonic particle in the Universe) this translates to an average of
\begin{align}\label{eq:n0}
    \langle n_{\rm H}\rangle_{0} \approx 2.5 \times 10^{-7} \ {\rm cm}^{-3} \ \rm{,}
\end{align}

\noindent as mentioned, a million times lower than the typical densities in the ISM. Given that the vast majority of the baryons are in the IGM (see Section~\ref{sec:hydro}), $\langle n_{\rm H}\rangle_{0}$ is comparable to the actual average IGM density. Considering the redshift evolution, we have
\begin{align}
    \langle n_{\rm H}\rangle = \langle n_{\rm H}\rangle_{0} \ (1+z)^3  \ \rm{,} \label{eq:evol1}
\end{align}

\noindent (similar equations hold for $\langle \rho \rangle$ and $\langle \rho_b \rangle$), i.e. the average number density of hydrogen was larger in the past because the Universe was smaller. For instance, at $z=3$, $\langle n_{\rm H}\rangle$ was $64$ times larger than $\langle n_{\rm H}\rangle_{0}$, i.e. $\langle n_{\rm H} (z=3)\rangle = 1.6\times 10^{-5}$\,cm$^{-3}$, but this is still a much lower value than the typical ISM densities.\footnote{As an exercise, the reader may compute the redshift where the $\langle n_{\rm H}\rangle$ was comparable to the typical densities in the ISM.} \sk

Assuming that baryons follow dark matter (and hence total matter), then 
\begin{align}
    \label{eq:cte_ratio}
    \rho_{\rm b}=\left(\frac{\Omega_b}{\Omega_m}\right)\,\rho \, , 
\end{align}
\noindent where $(\Omega_b/\Omega_m)= {\rm Const.} \approx0.17$ is the cosmological ratio of the baryon and matter density parameters \citep{planck2020}, and we can write the density contrast $\delta$\ in terms of just baryonic densities or just hydrogen number densities, i.e.,
\begin{align}\label{eq:overdensity3}
    \delta = \frac{\rho_{\rm b} - \langle \rho_{\rm b} \rangle}{\langle \rho_{\rm b} \rangle} = \frac{n_{\rm H} - \langle n_{\rm H} \rangle}{\langle n_{\rm H} \rangle} \, \rm{.}
\end{align}

\noindent In this Chapter we will refer to the over/under-density ratio (for $\delta>0$ and $\delta<0$, respectively) as,
\begin{align}
\label{eq:delta_mult}
    1+\delta = \frac{\rho}{\langle \rho \rangle}=\frac{\rho}{\rho_c}= \frac{\rho_b}{\langle \rho_b \rangle} = \frac{n_{\rm H}}{\langle n_{\rm H} \rangle}  \, \rm{.}
\end{align}

The assumption that baryons follow dark matter in a one-to-one correspondence should only hold in the general case that astrophysical processes (e.g. galaxy feedback) do not severely redistribute baryons. This is certainly not the case in the ISM, while in the CGM at $1+\delta \gtrsim 200$ (comparable to the virial radii of dark matter halos) this may or may not be true (given the uncertainties in our understanding of galaxy feedback processes; \citealt{tumlinson2017, amodeo2021,faucher2023}). 
But far away from galaxies at $1+\delta \lesssim 10-20$, this assumption should most likely hold.\footnote{The very close correspondence between the \lcdm\ predictions on these moderate overdensities scales, and the observed properties of the IGM traced by the \lya-forest (see Section~\ref{sec:qal_survey}) validates this assumption, at least at $z\gtrsim 2$ (see also Chapter~\chcosmolya).} These are the relevant overdensity scales for the IGM.\smallskip

Cosmic environments are defined by topological features like whether matter is distributed in halos (nodes), filaments, sheets, or voids, and studying them is currently a very active area of research (see Chapter~\chcosmicweb). Clearly, density contrast alone is not sufficient to capture the complexities of the multi-scale network and the non-linear physics involved in the growth of these structures (see Chapter~\chnonlinear). For this reason, in modern astrophysical studies of the IGM (and baryons in general) we depend on hydrodynamical simulations to provide more realistic scenarios to interpret observations. 

\subsection{Insights from hydrodynamical simulations}\sk
\label{sec:hydro}

\begin{figure}[t]
\centering
\includegraphics[width=0.95\textwidth]{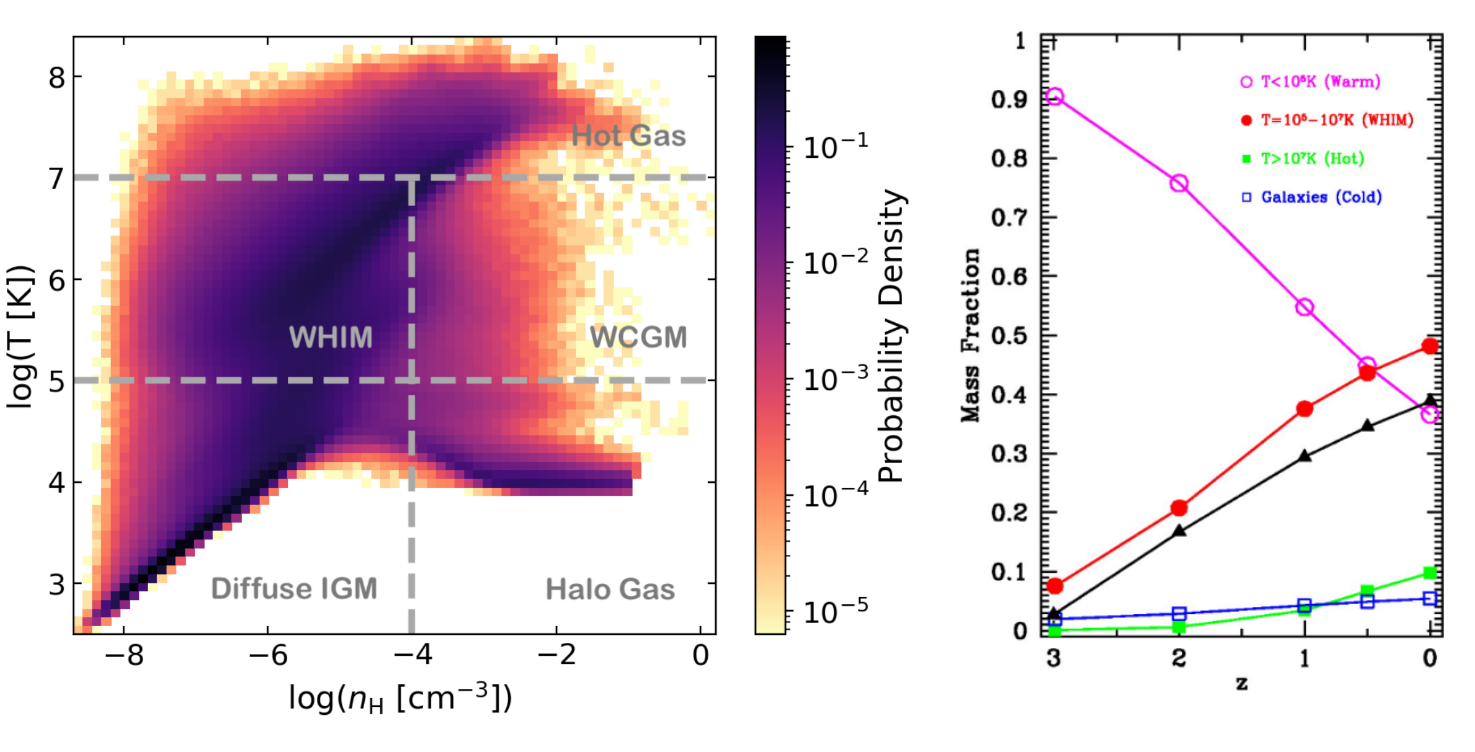}

\caption{Left: Density-temperature diagram for diffuse baryons at $z=0$ from a cosmological hydrodynamical simulation (Illustris TNG; \citealt{nelson2019}). The five phases presented in Section~\ref{sec:hydro} are labeled. Figure from \citet{galarraga2021}. Right: Mass fractions of the different gas phases as a function of redshift from the hydrodynamical simulation presented by \citet{cen2006}. Although the definition of the gas phases are slightly different than those presented in Section~\ref{sec:hydro}, this figure illustrates that the diffuse baryons (those outside galaxies) account for $\gtrsim 90$\,\% of the baryons in the Universe. At $z\gtrsim 2$, the dominant phase is the diffuse IGM (pink), whereas at $z\lesssim 0.5$ both the diffuse IGM and WHIM (red or black) dominate in almost equal proportion ($\sim 40$\,\%~each). The exact amount of baryons in the WHIM critically depends on galaxy feedback models. Here, a model with (red) and without (black) galactic winds are shown for reference. Figure adapted from \citet{cen2006}.}
\label{fig:phase_diagram}
\end{figure}

\subsubsection{Gas phases of diffuse baryons}\sk
In contrast to dark matter that only interacts gravitationally, the baryons in the IGM are also affected by pressure (collisions) and radiative processes. The advent of cosmological hydrodynamical simulations since the $1990$'s have helped to develop a fairly good understanding and intuition of the key physical processes that govern the different gas phases of the IGM \citep{cen1999,dave2001}. Despite all the underlying complexity, these simulations indicate that diffuse baryons reside in roughly $5$ gas phases based on densities and temperatures (see Figure~\ref{fig:phase_diagram}, left panel) defined as follows:\sk

\begin{itemize}
    \item {\bf Diffuse intergalactic medium (IGM):} at $n_{\rm H} \lesssim 10^{-4} (1+z)$\,\cc\ and $T<10^5$\,K, this phase dominate the baryon budget of the Universe. In terms of overdensities, these correspond to $1+ \delta \lesssim 100$.  At these temperatures and densities, this gas is mostly photoheated and photoionized, and a tight correlation between density and temperature arises (see Section~\ref{sec:thermal}). This phase is traced by the \lya-forest (see Section~\ref{sec:qal_class}; see also \citealt{rauch1998}).\sk
    
    \item {\bf Warm-hot intergalactic medium (WHIM):} at $n_{\rm H} \lesssim 10^{-4}(1+z)$\,\cc\ and $10^{5} \le T < 10^7$\,K, this phase builds up in importance as the Universe evolve and the large scale structures become more massive and dense. Most of this gas is shock heated either due to gravitational collapse and/or by galaxy feedback mechanisms, and resides in mild overdensities of $1 \lesssim 1+\delta \lesssim 100$ close to galaxies (e.g. cosmological filaments connecting galaxy clusters and/or galaxy clusters outskirts). At these temperatures, the gas is mostly collisionally ionized (see Section~\ref{sec:ionization}). This phase has been very challenging to observe directly (see Section~\ref{sec:qal_survey}). \sk
    
    \item {\bf Halo gas (cold):} at $ 10^{-4}(1+z) \lesssim n_{\rm H} \lesssim 10^{-1}$\,\cc\ and $T<10^{5}$\,K, this phase corresponds to `cool' gas within galaxy halos, including the CGM or diffuse ISM within galaxies, typically at $T\sim 10^4$\,K. At densities higher than $\approx 10^{-1}$\,\,\cc, the gas is typically associated with star-forming regions in the ISM, and thus not included here (but see Chapters~\chism\ and \chgasingal). In terms of overdensities, this gas is typically within or around virialized halos at $1+\delta \gtrsim 200-1000$. At these temperatures and densities, it is expected that the gas is mostly photoionized. This phase can be observed either in emission (for the densest parts) and/or in absorption against a background source. Most of the intervening low-ionization metal absorption lines observed towards quasars are tracers of this phase (e.g. \citealt{werk2014}; see Chapter~\chcgm). \sk
    
    \item {\bf Warm circumgalactic medim (WCGM):} at $ 10^{-4}(1+z) \lesssim n_{\rm H} \lesssim 10^{-1}$\,\cc\ and $10^{5} \le T < 10^7$\,K, this phase corresponds to `warm' gas within galaxy halos, typically in the CGM at overdensities $1+\delta \gtrsim 200-1000$. At these temperatures and densities, the gas is likely to be collisionally ionized (see Section~\ref{sec:ionization}). This phase can be observed in emission and/or absorption against a background source. Most of the intervening high-ionization metal absorption lines are likely tracing this phase (e.g. \citealt{aracil2004,tripp2008,burchett2019}; see Chapter~\chcgm).\sk 
    
    \item {\bf Hot gas:} at $T>10^7$\,K (regardless of density) this gas is tracing a `hot' medium, with temperatures comparable or larger than the virial temperature of massive galaxy clusters. This gas is mostly shock-heated and collisionally ionized. This phase can be observed in $X$-ray emission via Bremsstrahlung radiation (free-free emission) and highly ionized line emission, e.g. in the intracluster medium (see Chapter~\chclusters).\sk 
\end{itemize}

\noindent The exact density threshold adopted to divide these phases varies between different researchers and/or simulations, but it broadly corresponds to the IGM/CGM transition at the outskirts of virialized structures (galaxy halos). This limit can also vary as a function of redshift: for instance, the quoted value of $10^{-4}(1+z)$\,\cc\ is the one used by \citet{martizzi2019}, which captures this transition sufficiently well across various cosmic epochs for their analyzed simulation (but see \citealt{dave2010} for a different one). Similarly, star-formation requires a minimum density, and the limit of $0.13$\,\cc\ (or similar) is adopted as the interface between the diffuse medium and the star-forming gas (e.g. within the ISM). On the other hand, the temperature threshold at $T=10^5$\,K corresponds to the temperature where collisional ionization starts to become important for hydrogen (see Section~\ref{sec:ionization}), and is used by most researchers. Finally, the inclusion of the `hot gas' as a separate phase is not universally adopted, and some researchers divide these diffuse baryons just into the other four phases. \sk

\subsubsection{Baryon budget and redshift evolution}\sk

Out of these different phases, the diffuse IGM largely dominates the baryon budget at $z\gtrsim 2$, whereas at $z\lesssim 0.5$ both the WHIM and diffuse IGM become almost equally important (see Figure~\ref{fig:phase_diagram}, right panels). The diffuse IGM and the WHIM are the primary focus of this Chapter. We refer the reader to Chapters~\chcgm\, \chgasingal\ and~\chclusters\ (and references therein) for more details on the other three phases. \sk

\begin{figure}[t]
\centering
\includegraphics[width=1\textwidth]{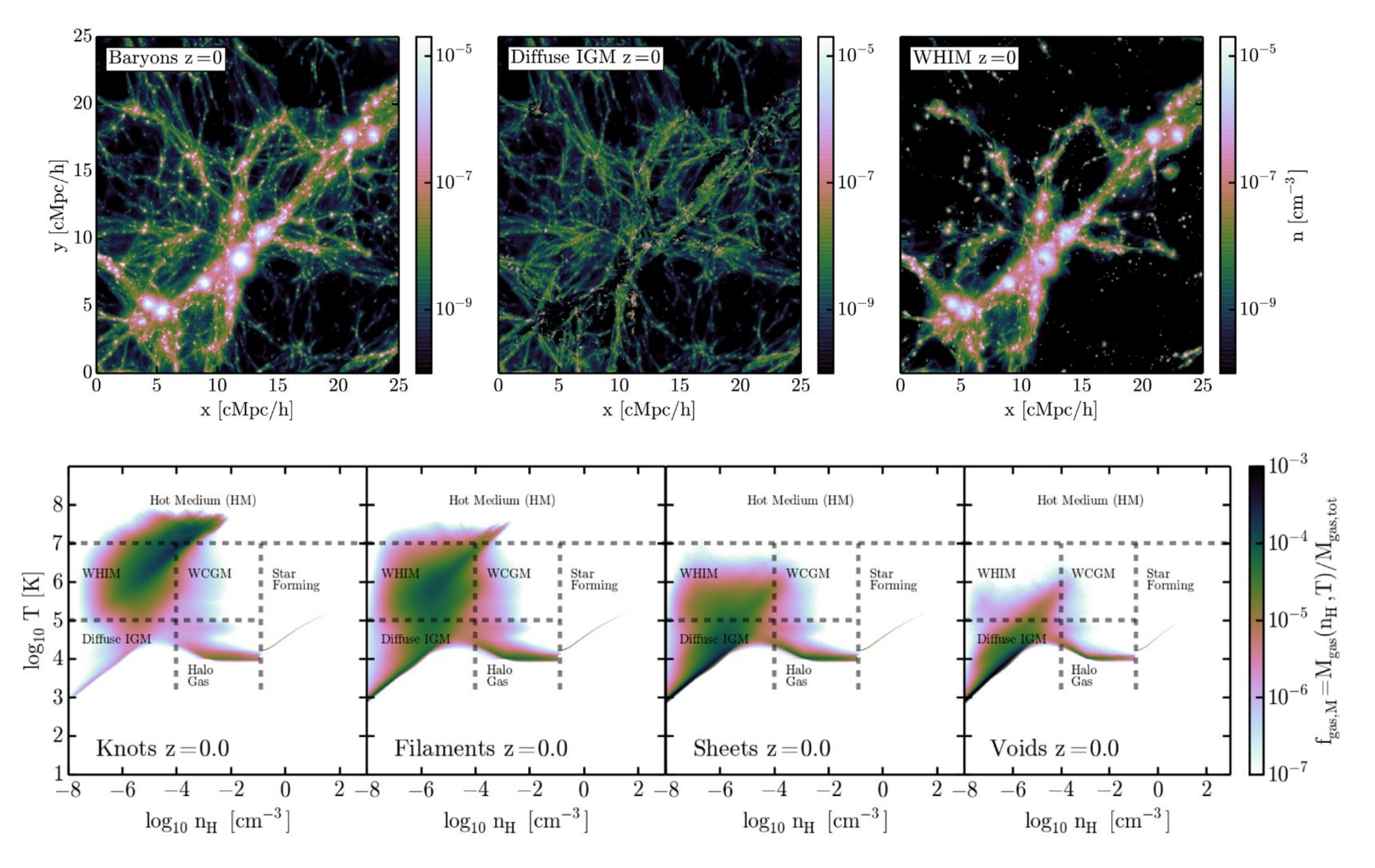}
\caption{Top: The $z=0$ distribution of all baryons (left), baryons in the diffuse IGM (middle) and baryons in the WHIM (right), over tens of co-moving Mpc (cMpc) from a state-of-the-art cosmologial hydrodynamical simulation as presented by \citet{martizzi2019}. The color bar represents density. Bottom: The density-temperature diagram for different cosmic environments (from left to right): knots, filaments, sheets and voids. The color bar represents total gass mass fractions. This figure illustrates that the diffuse IGM is mostly located in voids, sheets and filaments, while the WHIM is mostly found in knots and dense filaments, usually connecting galaxy clusters. Figure adapted from \citet{martizzi2019}.}
\label{fig:martizzi}
\end{figure}

Baryons can transit from one phase to another as a function of time as the Universe expands, cosmic structures grow, and galaxies evolve. Early studies from hydrodynamical simulations predicted the existence of a significant fraction ($\gtrsim 40-50$\,\%) of baryons in the WHIM residing in large and dense cosmological filaments at $z\lesssim 0.5$ \citep{cen1999, dave2001}. This WHIM has been very elusive to observe \citep{bregman2007, shull2012} but is currently our best candidate to host the so-called `missing baryons' \citep{fukugita1998}. At any given time, the relative importance of each phase also depends on the environment: theoretical studies have established that the diffuse IGM is mostly located in voids, sheets and filaments, while the WHIM is mostly found in dense filaments and knots (e.g. \citealt{martizzi2019}; see Figure~\ref{fig:martizzi}).\sk

Hydrodynamical simulations provide important insights regarding the complexities of the IGM in different environments, but let us come back to more basic physical principles that will allow us to model observations of the IGM.

\subsection{Modeling ionization and thermal state}\sk
\label{sec:thermal_ionization}

Both the ionization and thermal state of the IGM are regulated by the radiative and collisional processes that affect a gaseous medium. In particular, the properties of the IGM are influenced by processes such as: cosmic expansion, gravitational collapse, shock heating from large-scale structure formation, and energy input from quasars, hot stars, and supernovae.

\subsubsection{Ionization state}\sk
\label{sec:ionization}

The ionization state of a medium reflects the balance between ionizing sources and recombination processes. The IGM is predominantly ionized due to high-energy photons from quasars and star-forming galaxies that contribute to a UV background (UVB) radiation that permeates the Universe's volume. In this Section we will present an overview of the dominant mechanisms for determining the ionization state of the IGM, including photoionization, collisional ionization, radiative recombination, and ionization equilibrium.

\paragraph{Photoionization}\sk

Let us consider a gas composed of hydrogen atoms. If this gas is receiving incident radiation of sufficient energy, then the neutral atoms ($H^0$, also referred to as \hi) will interact with it and can become photoionized, by losing an electron: 
\begin{align}
\label{eq:photoion}
   H^0 + h\nu \rightarrow H^{+} + e^{-} \ \rm{,}
\end{align}

\noindent where $H^+$ represents ionized hydrogen (i.e. a free proton, also referred to as \hii), $h$ is the Planck constant, $\nu$ is the photon frequency, and $e^-$ represents a free electron. The rate at which \hii\ ions are produced by photoionization is:
\begin{align}
\label{eq:rate_pi}
  \frac{{\rm d}n_{\rm HII}}{{\rm d}t} = n_{\rm HI}n_{\gamma} \sigma_{\rm p.i.} c \, \rm{,}
\end{align}

\noindent where $n_{\rm HI}$ is the number density of neutral hydrogen, $n_{\gamma}$ is the number density of photons, $\sigma_{\rm p.i.}$ is the cross-section for photoionization,
and $c$ is the speed of light. A good approximation of the $\sigma_{\rm p.i.}$ for $h\nu>13.6$\,eV is given by \citet{draine2011}:
\begin{align}
  \sigma_{\rm p.i.} & \approx 0.225 a_0^2 \left(\frac{h\nu}{13.6\,{\rm eV}}\right)^{-3} \\
  & \approx 6.30 \times 10^{-18} \, {\rm cm}^2 \left(\frac{h\nu}{13.6\,{\rm eV}}\right)^{-3} \, \rm{,}
  \label{eq:sigma_pi}
\end{align}
\noindent with $a_0$ being the Bohr radius. The rapid decrease of $\sigma_{\rm p.i.}$ with frequency indicates that photons with energies just above $13.6$\,eV are the dominant ones in this process. Finally, the photoionization rate is defined as 
\begin{align}
 \Gamma_{\rm HI} \equiv  \int_{\nu_{\rm th}}^{\infty}n_{\gamma}\sigma_{\rm p.i.} c\, {\rm d}\nu = \int_{\nu_{\rm th}}^{\infty}\frac{u(\nu)}{h \nu}\sigma_{\rm p.i.} c\,{\rm d}\nu \, \rm{,}
\end{align}

\noindent where $\nu_{\rm th}$ is the photon frequency threshold that can ionize \hi\ ($\nu_{\rm th} = 13.6\, {\rm eV}/h=3.2 \times 10^{15}$\,Hz, corresponding to a wavelength of $912$\,\AA), and $u(\nu)$ is the specific energy density of the radiation field ($n_{\gamma} = \frac{u(\nu)}{h\nu}$), such that Equation~\ref{eq:rate_pi} can be written as:
\begin{align}
  \frac{{\rm d}n_{\rm HII}}{\rm{d}t} = n_{\rm HI}  \Gamma_{\rm HI} \, \rm{.}
\end{align}

\begin{figure}[t]
\centering
\includegraphics[width=0.65\textwidth]{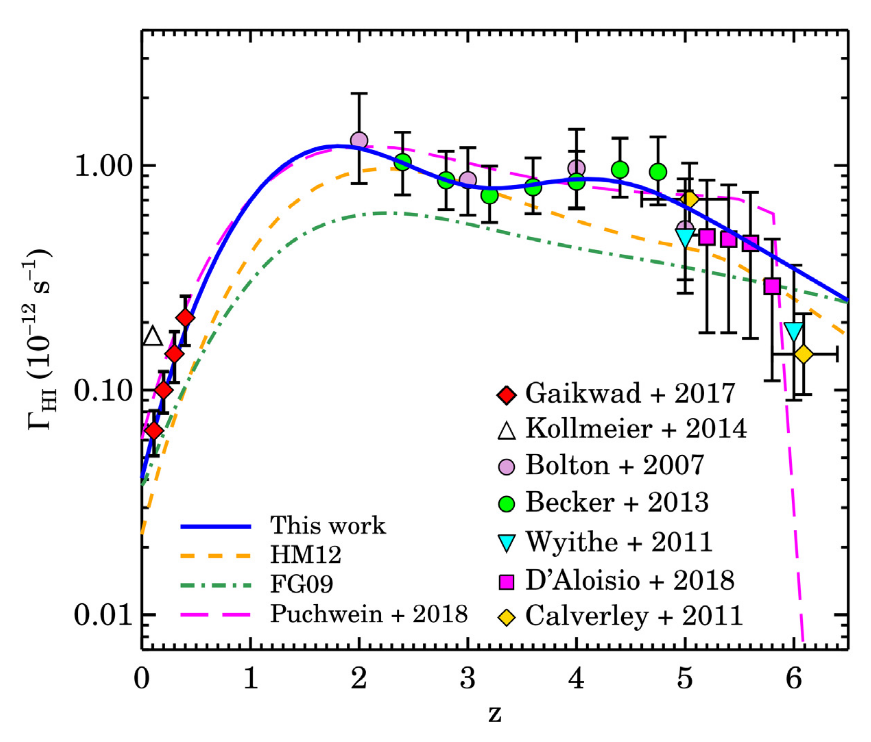}
\caption{Models and observational constraints for the IGM photoionization rates from the UVB ($\Gamma_{\rm HI}$) as a function of redshift. The models from \citet{khaire2019} (blue-solid line) and from \citet{haardt2012} (yellow-dashed line), among others (see references in \citealt{khaire2019}), are shown. Figure from \citet{khaire2019}.}
\label{fig:ionization}
\end{figure}

The photoionization rate corresponds to the rate at which a single neutral hydrogen atom is ionized, usually expressed in units of s$^{-1}$. Because the IGM is a highly ionized optically thin medium, these high-energy photons (after escaping their original galaxies) travel freely contributing to the UVB responsible for maintaining such an ionization state. Figure~\ref{fig:ionization} shows models and observational constraints of $\Gamma_{\rm HI}$ as a function of redshift. Estimated values for $\Gamma_{\rm HI}$ from the UVB range between $\Gamma_{\rm HI} \sim 10^{-12}$\,s$^{-1}$ at $z>2$ and $\Gamma_{\rm HI} \sim 1-10 \times 10^{-14}$\,s$^{-1}$ at $z\sim 0$ (e.g. \citealt{haardt2012, khaire2019}). The photoionization timescale, defined as $t_{\rm p.i.} \sim \frac{1}{\Gamma_{\rm HI}}$, is then $\sim 40\,000$\,yr at $z=2-5$ and $\sim 0.4-4$\,Myr at $z\lesssim 0.5$.


\paragraph{Collisional ionization}\sk

If the temperatures of the gas is sufficiently high, then interactions with free electrons can also collisionaly ionized the neutral atoms:
\begin{align}
   H^0 + e^- \rightarrow H^{+} + 2e^{-} \, \rm{,}
\end{align}

\noindent and the rate at which ions $H^+$ are produced by collisional ionization is:
\begin{align}
\label{eq:ci_rate}
  \frac{\rm{d}n_{\rm HII}}{\rm{d}t} = n_{\rm HI}n_{e}\langle\sigma_{\rm c.i.}v\rangle  \, ,
\end{align}

\noindent where $n_{e}$ is the number density of free electrons, $\sigma_{\rm c.i.}$ is the cross-section for collisional ionization, and $v$ is the relative speed of free electrons with respect to the neutral hydrogen atoms. Given that electrons are much lighter than \hi\ atoms, $v$ can be approximated by the electron speeds. The term $\langle \sigma_{\rm c.i.} v \rangle$ represents the average product of $\sigma_{\rm c.i.}$ and $v$ over the distribution of speeds (or energies) given by the gas temperature $T$. Usually, this $\langle \sigma_{\rm c.i.} v \rangle$ term is referred to as the collisional ionization rate coefficient, $k_{\rm c.i.}(T)$, such that Equation~\ref{eq:ci_rate} can be written as:
\begin{align}
  \frac{\rm{d}n_{\rm HII}}{\rm{d}t} = n_{\rm HI}n_{e}k_{\rm c.i.}(T)  \, ,
\end{align}

\noindent Assuming kinetic equilibrium, $k_{\rm c.i.}(T)$, can be computed from a Maxwellian distribution of energies (given the speeds) as,
\begin{align}
\label{eq:kci_maxwell}
  k_{\rm c.i.}(T) = \left(\frac{8kT}{\pi m_{e}}\right)^{1/2} \int_{I_H}^{\infty} \sigma_{\rm c.i.} \frac{E}{kT} \exp{\left(-\frac{E}{kT}\right)} \frac{{\rm d}E}{kT}  \, ,
\end{align}

\noindent where $m_{e}$ is the electron mass, $k$ is the Boltzmann constant, and $I_H$ is the energy threshold for ionizing \hi\ ($I_H=13.6$\,eV). The precise values for $\sigma_{\rm c.i.}$ are determined from experiments, but a useful approximation for energies $E>I_H$ is given by \citep{draine2011}:
\begin{align}
    \sigma_{\rm c.i.} &\approx 1.07\pi a_0^2\left(1-\frac{I_H}{E}\right)  \, ,
\end{align}

\noindent from which Equation~\ref{eq:kci_maxwell} yields,
\begin{align}
\label{eq:kci_approx}
  k_{\rm c.i.}(T) &\approx 1.07\pi a_0^2 \left(\frac{8kT}{\pi m_{e}}\right)^{1/2} \exp{\left(-\frac{I_H}{kT}\right)} \\
  & \approx   5.85 \times 10^{-9} \, {\rm cm}^3\,{\rm s}^{-1} \, \left(\frac{T}{10^4\,{\rm K}}\right)^{1/2}\exp{\left(-\frac{I_H}{kT}\right)} \, . 
  \label{eq:kci_approx2}
\end{align}

\noindent The exponential cut-off of $k_{\rm c.i.}$ at low kinetic energies, imply that collisional ionization is negligible at temperatures $T\ll I_{H}/k =1.58\times 10^5$\,K, as expected. Therefore, for applications within the context of the diffuse IGM this mechanisms can be ignored, but it is highly relevant for the WHIM.\sk

\paragraph{Radiative recombination}\smallskip 

The opposite process to ionization is recombination. In the case of radiative recombination, the ionized hydrogen captures a free electron and a photon is emitted:
\begin{align}
\label{eq:recomb}
   H^{+} + e^{-} \rightarrow H^0 + h\nu   \, .
\end{align}

\noindent The rate at which \hi\ atoms are produced in this case is:
\begin{align}
\label{eq:rate_rr}
  \frac{{\rm d}n_{\rm HI}}{{\rm d}t} = n_{\rm HII}n_{e} \langle \sigma_{\rm r.r.} v \rangle  \, ,
\end{align}

\noindent where $n_{\rm HII}$ is the number density of ionized hydrogen, $n_{e}$ is the number density of free electrons, $\sigma_{\rm r.r.}$ is the cross-section for the radiative recombination, and $v$ is the relative speed of free electrons with respect to the ionized hydrogen. Again, given that electrons are much lighter than protons, $v$ can be approximated by the electron speeds. Similarly to the collisional ionization case, the term $\langle \sigma_{\rm r.r.} v \rangle$ represents the average product of $\sigma_{\rm r.r.}$ and $v$ over the distribution of speeds (or energies) given by the gas temperature $T$, and  is usually referred to as the radiative recombination rate coefficient, $\alpha(T)$. With this notation, Equation~\ref{eq:rate_rr} can be written as:
\begin{align}
  \frac{{\rm d}n_{\rm HI}}{{\rm d}t} = n_{\rm HII}n_{e} \alpha(T)  \, .
\end{align}

\noindent Again, assuming kinetic equilibrium, $\alpha(T)$, can also be computed from a Maxwellian distribution of energies, but we need to consider all possible quantum numbers ($n,l$) of the hydrogen atom to which electrons can be recombined. For a particular particular ($n,l$) quantum number, we have:
\begin{align}
\label{eq:alfa_maxwell}
  \alpha_{nl}(T) = \left(\frac{8kT}{\pi m_e}\right)^{1/2} \int_0^{\infty} \sigma_{{\rm r.r.},nl} \frac{E}{kT}\exp \left(-\frac{E}{kT}\right) \frac{{\rm d}E}{kT} \, ,
\end{align}

\noindent and the effective $\alpha(T)$ is obtained by summing over all possible quantum numbers. As pointed out by \citet{baker1938}, two extreme cases could be considered: i) an optically thin cloud where all photons produced by recombination escape the system (Case A), and ii) an optically thick cloud where photons produced by recombination to the ground level ($n=1$) cannot escape because these are immediately reabsorbed (Case B). Thus we have:
\begin{align}
 \alpha_A(T) &= \sum_{n=1}^{\infty}\sum_{l=0}^{n-1} \alpha_{nl}(T)  \, , \\
 \alpha_B(T) &= \sum_{n=2}^{\infty}\sum_{l=0}^{n-1} \alpha_{nl}(T) =   \alpha_A(T) - \alpha_{1s}(T) \, ,
\end{align}
\noindent respectively. The exact values are determined by quantum mechanics and/or experiments, but these two cases can be approximated by \citep{draine2011}:
\begin{align}
\label{eq:case_A}
 \alpha_A(T) \approx 4.13 \times 10^{-13} \, {\rm cm}^3\,{\rm s}^{-1} \left(\frac{T}{10^4\,{\rm K}}\right)^{-0.71}   \, ,
\end{align}
\begin{align}
 \alpha_B(T) \approx 2.54 \times 10^{-13} \, {\rm cm}^3\,{\rm s}^{-1} \left(\frac{T}{10^4\,{\rm K}}\right)^{-0.81} \, ,
\end{align}

\noindent respectively. For applications regarding the IGM, we will be mostly concerned with Case A. The radiative recombination timescale is defined as $t_{\rm r.r.} \sim \frac{1}{n_e\alpha(T)}$, and thus depends on both density of free electrons and temperature. For the IGM, and parametrizing the electron density in terms of the mean density of the Universe, this corresponds to,
\begin{align}
    t_{\rm r.r.}\approx 9.7 \times 10^{18}\, {\rm s} \, \frac{\langle n_{\rm H}\rangle_0(1+z)^{-3}}{n_e}\left(\frac{T}{10^4\,{\rm K}}\right)^{0.71} \, ,
\end{align}
\noindent implying that for highly ionized gas at mean density, $t_{\rm r.r.} \gg t_{\rm p.i}$ at $z=0-6$. This explains why after reionization, most of the IGM remains highly ionized.

\paragraph{Ionization equilibrium and ionization fractions}\sk

A particularly interesting case is that of ionization equilibrium, in which the rates at which \hi\ are converted into \hii, and vice-versa, are equal:
\begin{align}
    \frac{{\rm d}n_{\rm HII}}{\rm{d}t} & = \frac{{\rm d}n_{\rm HI}}{\rm{d}t}\\    
    n_{\rm HI}  \Gamma_{\rm HI} + n_{\rm HI} n_{\rm HII}k_{\rm c.i.}(T) & = n_{\rm HII}^2 \alpha(T)
    \label{eq:ion_equi}
\end{align}

\noindent considering that $n_{e}=n_{\rm HII}$ for an overall neutral plasma. Let us define the neutral and ionized gas fractions as
\begin{align}
    x_{\rm HI} & \equiv \frac{n_{\rm HI}}{n_{\rm H}} = \frac{n_{\rm HI}}{n_{\rm HI}+ n_{\rm HII}} \, , \\
    x_{\rm HII} & \equiv \frac{n_{\rm HII}}{n_{\rm H}} = \frac{n_{\rm HII}}{n_{\rm HI}+ n_{\rm HII}} \, ,
\end{align}

\noindent respectively. Then, $n_{\rm HII} = x_{\rm HII}n_{\rm H}$ and $n_{\rm HI} = x_{\rm HI}n_{\rm H} = (1-x_{\rm HII})n_{\rm H}$. We can write Equation~\ref{eq:ion_equi} as a quadratic equation for $x_{\rm HII}$ as:
\begin{align}
\label{eq:ion_fraction}
    x_{\rm HII}^2 n_{\rm H}\{\alpha(T)+k_{\rm c.i.}(T)\} + x_{\rm HII} \{\Gamma_{\rm HI} - n_{\rm H}k_{\rm c.i.}(T)\} - \Gamma_{\rm HI} = 0 \, .
\end{align}

\noindent Solving this equation for a given $n_H$, $\alpha(T)$, $\Gamma_{\rm HI}$ and $k_{\rm c.i.}(T)$ provides the ionized fraction $x_{\rm HII}$ (and hence the neutral fraction $x_{\rm HI} = 1 - x_{\rm HII}$) of the gas. However, this solution is not necessarily straightforward, as it requires the knowledge of the thermal state ($T$) and shape and intensity of the radiation field encoded in $\Gamma_{\rm HI}$ for a given $n_H$. We note here that the thermal state of the gas is critically regulated by the amount of metal ions via radiative cooling and thus the metallicity of the gas is an important parameter to consider, especially in high density regions close to galaxies (see Section~\ref{sec:thermal}). Similarly, the sources and the interaction of the radiation field with the gas itself is also important (especially in dense environments), where radiative transfer equations must be explicitly solved for a given geometry. The inclusion of all these effects makes this problem sufficiently complex that dedicated software and/or the use of hydrodynamical simulations are required. For instance, the code \cloudy\ \citep{ferland1998} is widely used to infer the underlying physical conditions of the gas in widely diverse gaseous astrophysical environments. Although most complications arise in dense environments close to or within galaxies, some limiting cases can become good approximations in practice: \sk
\begin{itemize}
    \item {\bf Photoionization equilibrium (PIE):} Assuming temperatures sufficiently low that $k_{\rm c.i.}$ can be ignored ($T<10^5$\,K; see Equation~\ref{eq:kci_approx2}), then Equation~\ref{eq:ion_fraction} is approximated to:
\begin{align}
    x_{\rm HII}^2 \frac{n_{\rm H}\alpha(T)}{\Gamma_{\rm HI}} + x_{\rm HII}  - 1 \approx 0 \, ,
\end{align}

\noindent which can be easily solved.\footnote{For a positive solution we have, $x_{\rm HII} = \frac{-1 + \sqrt{1+4a}}{2a}$, with $a\equiv n_{\rm H}\alpha(T)/\Gamma_{\rm HI}$.} Here, let us consider two limits based on comparing $n_{\rm H}$ with the ratio $\Gamma_{\rm HI}/\alpha(T)$. When $n_{\rm H} \ll \Gamma_{\rm HI}/\alpha(T)$ (sufficiently low density environments) then,
\begin{align}
\label{eq:ion_frac}
    x_{\rm HII} \approx 1 - n_{\rm H} \alpha(T)/\Gamma_{\rm HI} \, ,
\end{align}
\noindent implying a neutral fraction of
\begin{align}
\label{eq:neutral_frac}
    x_{\rm HI} \approx n_{\rm H} \alpha(T)/\Gamma_{\rm HI} \, .
\end{align}
\noindent In the case of the IGM at $T\sim 10^4$\,K, $\Gamma_{\rm HI}/\alpha(T) \approx 0.2 - 2$\,\cc\ (depending on redshift) and $n_{\rm H}\lesssim 10^{-4}$\,\cc, and thus Equations~\ref{eq:ion_frac}~and~\ref{eq:neutral_frac} are actually very good approximations. On the other hand, for the case $n_{\rm H} \gg \Gamma_{\rm HI}/\alpha(T)$ (sufficiently high density environments) then,
\begin{align}
\label{eq:ion_frac_2}
    x_{\rm HII} &\approx \sqrt{\Gamma_{\rm HI}/n_{\rm H} \alpha(T)} \, , \\
     x_{\rm HI} &\approx 1 - \sqrt{\Gamma_{\rm HI}/n_{\rm H} \alpha(T)} \, ,
\label{eq:neutral_frac_2}
\end{align}
\noindent which are more appropriate for some ISM and/or some CGM environments.\sk

\item {\bf Collisional ionization equilibrium (CIE):} Assuming temperatures sufficiently high that collisional ionization dominates ($T>10^5$\,K) and that $\Gamma_{\rm HI}$ can be ignored, then Equation~\ref{eq:ion_fraction} implies that the ionized and neutral fractions become independent of density and are only functions of temperature,
\begin{align}
\label{eq:CIE}
    x_{\rm HII} &\approx \frac{k_{\rm c.i.}(T)}{ \alpha(T)+k_{\rm c.i.}(T)} \, , \\
     x_{\rm HI} &\approx \frac{\alpha(T)}{ \alpha(T)+k_{\rm c.i.}(T)} \, .
\end{align}
\noindent This approximation is relevant for the WHIM, WCGM and hot gas environments.

\end{itemize}

\subsubsection{Thermal state}\sk
\label{sec:thermal}
The thermal state of the IGM is also regulated by collisional and radiative processes, and is usually estimated by balancing the heating and cooling rates in the context of an expanding Universe. In this Section, we will describe the most important mechanisms and modeling for determining the thermal state of the IGM.\sk

\paragraph{Heating mechanisms}\sk

The primary sources of heating in the IGM are: \sk
 
\begin{itemize}
    \item {\bf Photoheating:} produced by photons that can ionize either hydrogen or helium. Once an atom of hydrogen or helium is ionized, the free electron(s) produced by the interaction (e.g. Equation~\ref{eq:photoion}) carry part of the incident photon energy as kinetic energy. Collisions between these free electrons and the rest of ions heat the medium. Only photons with $E>h\nu_{\rm th}^{\rm HI}=13.6$\,eV can ionize hydrogen, whereas for partially or for fully ionizing helium, photons with $E>h\nu_{\rm th}^{\rm HeI}=24.6$\,eV and $E>h\nu_{\rm th}^{\rm HeII}=54.4$\,eV are required, respectively. The photoheating rate is then:
    \begin{align}
        \mathcal{H}_{\rm photo}= n_{\rm HI} \int_{\nu_{\rm th}^{\rm HI}}^{\infty}\frac{u(\nu)}{h \nu}\sigma_{\rm p.i.}^{\rm HI} c \, (h\nu - h\nu_{\rm th}^{\rm HI})\,{\rm d}\nu + n_{\rm HeI} \int_{\nu_{\rm th}^{\rm HeI}}^{\infty}\frac{u(\nu)}{h \nu}\sigma_{\rm p.i.}^{\rm HeI} c \, (h\nu - h\nu_{\rm th}^{\rm HeI})\,{\rm d}\nu + n_{\rm HeII} \int_{\nu_{\rm th}^{\rm HeII}}^{\infty}\frac{u(\nu)}{h \nu}\sigma_{\rm p.i.}^{\rm HeII} c \, (h\nu - h\nu_{\rm th}^{\rm HeII})\,{\rm d}\nu \, ,
    \end{align}
    \noindent where $n_{\rm HI}$, $n_{\rm HeI}$ and $n_{\rm HeII}$ are the corresponding densities, and $\sigma_{\rm p.i.}^{\rm HI}$, $\sigma_{\rm p.i.}^{\rm HeI}$, $\sigma_{\rm p.i.}^{\rm HeII}$ are the corresponding photoionization cross-sections. The sources of these ionizing photons (i.e. the $u(\nu)$ radiation field) are thought to be produced within galaxies, either from young massive stars (dominating at early cosmic times) and/or quasars (dominating at late cosmic times and at $E>54.4$\,eV); however, the mechanism that allows these high-energy photons escape the galaxies remains quite uncertain.\footnote{Note that high-energy photons ($E > 13.6$\,eV) have large cross-sections (Equation~\ref{eq:sigma_pi}) and thus are likely to be absorbed by the surrounding gas and thus remain within the galaxies. Similarly, interaction of these photons with dust grains in the ISM also prevent them from easily escaping.}\sk

    \item  {\bf Adiabatic heating:} produced by the accretion and collapse of matter due to gravity. As baryons fall into the potential wells of overdense structures (e.g. sheets, filaments, nodes or halos) they gain kinetic energy adiabatically.\footnote{\label{fn}This process is a consequence of the first law of thermodynamics which states that, in the absence of heat exchange with the surroundings, the compression/expansion of the system will lead to an increase/decrease in temperature.}\sk

    \item {\bf Shock heating from structure formation:} also produced by the accretion and collapse of matter due to gravity. In contrast to adiabatic heating, here the kinetic energy is added via shocks. When accretion speeds are larger than the sound speed of the medium, then a shock front is produced and the gas is suddenly heated due to compression. Shock heating is responsible for the majority of the gas in the WHIM and the hot gas phases of IGM, especially within large and dense cosmological filaments and galaxy clusters (e.g. the ICM). \sk
   
    \item {\bf Shock heating from galaxy feedback processes:} produced by energetic outflows from galaxies, driven primarily by stellar winds and supernovae (SNe) from star-forming regions and/or AGN activity. These feedback processes can eject large amounts of gas at high velocities (depending on the models), creating localized shocks as the outflows collide with the surrounding IGM. These shocks can heat the gas to $T>10^5$\,K thus redistributing baryons to the WHIM, warm CGM and hot gas phases. However, the actual relative importance of this process is not well established, as it critically depends on the treatment and modeling used in simulations for star-formation and AGN, often done at the sub-grid level (unresolved; see \citealt{somerville2015} for a review).\sk

\end{itemize}
\paragraph{Cooling mechanisms}\sk
\noindent The primary mechanisms of cooling the gas in the IGM are:\sk

\begin{itemize}

     \item {\bf Adiabatic cooling}: produced by the expansion of the Universe. As the density of non-virialized structures decreases, baryons lose kinetic energy adiabatically.\footnotemark[8] This mechanism is particularly important after EoR and dominates the cooling of the IGM at $z\lesssim 6$. \sk

    \item {\bf Compton cooling:} produced by free electrons that interact with lower-energy photons in the medium via scattering. This interaction transfers energy from the electrons to the photons, effectively cooling the medium. In the IGM, these low-energy photons are from the cosmic microwave background CMB), and thus the Compton cooling rate is \citep{weymann1965},
    \begin{align}
        \Lambda_{\rm Compt.} = 4n_ec\frac{\sigma_{\rm T}}{m_ec^2}a_{\rm rad}T^4_{\rm CMB}(z) \, k \, (T-T_{\rm CMB}(z)) \, ,
    \end{align}
    \noindent where $\sigma_{\rm T}$ is the Thomson cross-section, $a_{\rm rad}$ is the radiation density constant, and $T_{\rm CMB}(z)=T_{\rm CMB,0}(1+z)$ is the temperature of the CMB as a function of $z$, with $T_{\rm CMB,0}\approx 2.7$\,K. Compton cooling is more relevant for gas at $T \gtrsim 10^7$\,K, and also particularly important at $z\gtrsim 6$, i.e. before EoR (see Chapter~\cheor).\sk

    \item {\bf Collisional excitation cooling:} produced by free electrons that collide with ions and have enough energy to excite them. Once the ions return to their ground state by emitting photons, these carry energy away from the medium, thus cooling the gas. This process is more relevant for enriched gas at $10^4 < T \lesssim 10^6$\,K.\sk
    
    \item {\bf Collisional ionization cooling:} produced by free electrons that collide with ions and have enough energy to ionize them. The energy required to ionize the ions is effectively taken from the kinetic energy of the colliding electrons, thus cooling the gas. This process is more relevant for enriched gas at $T > 10^5$\,K.\sk 

    \item {\bf Radiative recombination cooling}: produced by free electrons that recombine into ions. The emitted photons in the process (Equation~\ref{eq:recomb}), carry energy away from the medium, thus cooling the gas. This process is more relevant for enriched gas at $10^4 < T \lesssim 10^5$\,K.\sk
    
    \item {\bf Bremsstrahlung (free-free emission) cooling:} produced by free electrons when these are deflected by charged ions (typically protons). Due to the acceleration of the electrons, photons are emitted which carry energy away from the medium. At the densities considered here, this process is more relevant for gas at $10^6 < T \lesssim 10^8$\,K, e.g. in galaxy clusters or massive groups (see Chapter~\chclusters).\sk 

\end{itemize}

\noindent Although in this Chapter we have mostly focused in hydrogen and helium, here it is important to emphasize that for collisional and radiative recombination cooling, the metallicity of the gas must be critically taken into account (specially in the densest and more enriched gas phases, like the CGM; see Chapter~\chcgm). Metal ions provide additional energy levels and transitions that allow the gas to lose energy more efficiently, especially at $10^{5}\lesssim T \lesssim 10^7$\,K where hydrogen and helium alone are not quite effective coolants \citep[e.g.][]{wiersma2009}. \sk

\begin{figure}[t]
\centering
\includegraphics[width=0.7\textwidth]{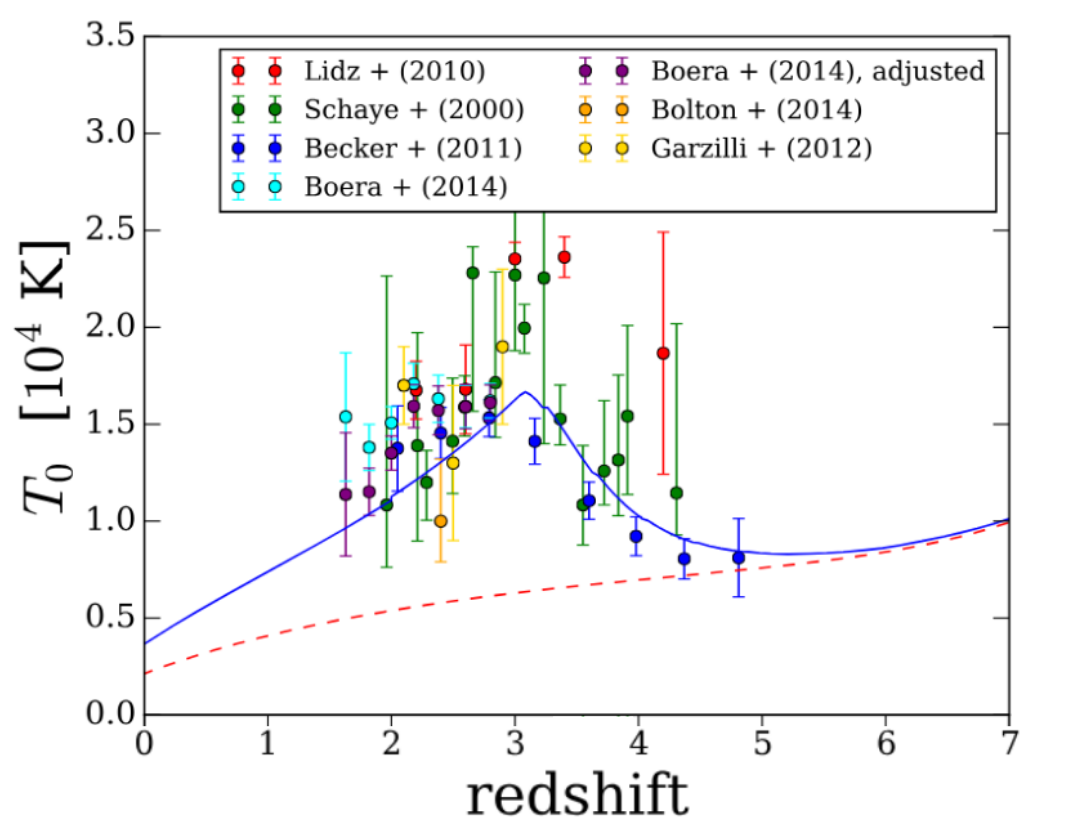}
\caption{Models and observations for the temperature of the IGM at mean density, $T_0$, as a function of redshift. Two models from \citet{upton2016} are shown: with (blue-solid line) and without (red-dashed line) \heii\ reionization. Observational constraints from several studies are also shown (see compilation by \citealt{upton2016}). Figure adapted from \citet{upton2016}.}
\label{fig:temperature}
\end{figure}

\paragraph{Modeling the thermal evolution of the IGM}\sk

The temperature evolution of the IGM is modeled as \citep{hui1997, upton2016},
\begin{align}
\label{eq:temp_evol}
    \frac{{\rm d}T}{{\rm d}t} = -2HT + \frac{2}{3}\frac{T}{(1+\delta)}\frac{{\rm d}\delta}{{\rm d}t} - \frac{T}{n_{\rm tot}}\frac{{\rm d}n_{\rm tot}}{{\rm d}t} + \frac{2}{3kn_{\rm tot}} \frac{{\rm d}Q}{{\rm d}t} \, ,
\end{align}

\noindent where ${\rm d}/{\rm d}t$ is the Lagrangian time derivative, $H$ is the Hubble parameter\footnote{For a flat \lcdm\ cosmology this is defined as $H\equiv H_0\sqrt{\Omega_{\rm m}(1+z)^3+\Omega_{\rm \Lambda}}$ where $H_0$ is the Hubble constant.}, $\delta$ is the density contrast, $n_{\rm tot}$ is the total number density of free baryonic particles (ions and electrons)\footnote{Despite electrons being leptons, these are usually included as `baryons' by astronomers.}, $k$ is the Boltzmann constant, and ${\rm d}Q/{\rm d}t$ is the heat gain per unit volume. The first two terms  correspond to adiabatic cooling and heating, respectively. The third term corresponds to the change of internal energy per particle as the number of particles vary. Typically this term is unimportant (${\rm d}n_{\rm tot}/{\rm d}t\approx 0$) except right after hydrogen and helium reionization. The fourth term includes all the other sources of heating and cooling previously mentioned, although in most modelings shock heating is treated separately or is not considered.\sk

This modeling has been successful at reproducing observational constraints at $z>1$ (without shocks). Figure~\ref{fig:temperature} shows two models for the evolution of the temperature of the IGM at mean density, $T_0$, as a function of redshift from \citet{upton2016}: with (blue line) and without (red dashed line) \heii\ reionization. Some measurements from various works using a variety of techniques (see references in \citealt{upton2016}) are also shown for comparison. The temperature of the IGM evolves slowly with redshift, with typical values $T_0\sim 1-2 \times 10^4$\,K. The peak of $T_0$ at $z\sim 3$ favors the interpretation of a late \heii\ reionization. These measurements and modeling also indicate that for the diffuse IGM between $2\lesssim z \lesssim 6$, photoheating from the UVB is the dominant source of heating, whereas both adiabatic and Compton cooling are the dominant coolants \citep{mcquinn2016b}. Therefore, for most applications concerning the diffuse IGM, ${\rm d}Q/{\rm d}t$, can be well approximated as
\begin{align}
 \frac{{\rm d}Q}{{\rm d}t} \approx  \mathcal{H}_{\rm photo} - \Lambda_{\rm Compt.} \, .
\end{align}


\paragraph{The density-temperature relation for the diffuse IGM}\smallskip

The diffuse IGM presents an strikingly tight correlation between density and temperature. This behavior is apparent in the temperature-density phase diagram of hydrodynamical simulations, such as those presented in Figures~\ref{fig:phase_diagram}~and~\ref{fig:martizzi}. The previous modeling of the thermal history (Equation~\ref{eq:temp_evol}) have helped to establish the origin of this remarkable feature \citep{miralda-escude1994, hui1997, theuns1998}. The ``observed''\footnote{Observed in hydrodynamical simulations.} relation has a power-law form,
\begin{align}
\label{eq:Tdensity}
    T \approx T_0 (1+\delta)^{\gamma -1} \, ,
\end{align}

\noindent with $\gamma>1$ and where $\gamma-1\approx 0.6$. Interestingly, the vast majority of the diffuse IGM is expected to follow this trend, starting right after re-ionization ends. Given that the photoheating timescales ($t=1/\Gamma_{\rm HI}$; see Section~\ref{sec:ionization}) are much shorter than the Hubble times, an equilibrium temperature is rapidly reached for a given density $n_{\rm H}$ (\citealt{hui1997}, see also \citealt{mcquinn2016b}). Assuming PIE in the low density regime (Equation~\ref{eq:neutral_frac}), $n_{\rm HI}\Gamma_{\rm HI} = n^2_{\rm H}\alpha(T)$ with $\alpha(T) \propto T^{-\beta}$ (see Section~\ref{sec:ionization}). Parametrizing $n_{\rm H}$ and $T$ as a function of overdensity $(1+\delta)$, Equation~\ref{eq:temp_evol} can be solved for a $T(1+\delta)$. Ignoring non-adiabatic cooling terms, this approach leads to a solution of the form $T = T_0(1+\delta)^{1/(1+\beta)}$ with $\beta\approx 0.7$ (Case A; see Equation~\ref{eq:case_A}), corresponding to $T= T_0(1+\delta)^{1/1.7} \approx T_0(1+\delta)^{0.6}$ \citep{miralda-escude1994, hui1997, theuns1998}. The previous argument indicates that the temperature-density relation arises from the balance between photo-heating and adiabatic cooling. However, \citet{mcquinn2016b} have argued that the slope is actually set by the fact that the ratio between the photo-heating rate and $n_{\rm H}^{2/3}$ is almost constant (see their equation~12), whereas adiabatic (and Compton) cooling only help the gas to reach this equilibrium faster.\sk

The previous treatment ignores the effect of shocks in the IGM. At $T\gtrsim 10^5$\,K, shock heating and collisions are important and Equation~\ref{eq:Tdensity} no longer holds. Indeed, shocks will tend to move gas away from any density-temperature relation, and thus these will increase the scatter around it if such a relation exists \citep{mcquinn2016b}. Shocks in the IGM become more important as the Universe evolves, especially at $z\lesssim 1$ (see Section~\ref{sec:hydro}). In the WHIM, shocks produced by structure formation are expected to dominate the heating, but galaxy feedback shocks may also be important (depending on the models). Although a similar temperature-density relation could be seen in the WHIM of some hydrodynamical simulations (e.g. \citealt{galarraga2021}; see Figure~\ref{fig:phase_diagram}), these are not as tight as the one in the diffuse IGM, and the slope is not set to a universal value (model dependent).

\subsection{Hydrodynamical equilibrium in the diffuse IGM}\sk
\label{sec:equilibrium}

Observations of the IGM are typically achieved via the absorption lines imprinted in the spectra of background quasars, and this technique can only access the neutral hydrogen atoms (see Section~\ref{sec:qal}). Thus, it is worthwhile finding an expression for the characteristic neutral hydrogen column density, $N_{\rm HI}$, for a given $n_{\rm H}$, $T$ and $\Gamma_{\rm HI}$, in this highly ionized medium.\sk

The general properties of the diffuse IGM can be understood based on stability arguments. Assuming IGM ``clouds'' are self gravitating systems in hydrostatic equilibrium, then the dynamical ($t_{\rm dyn} \sim 1/\sqrt{\rho G}$) and sound crossing time ($t_{\rm sc} \sim L/c_{\rm s}$) should be of the same order, $t_{\rm dyn} \sim t_{\rm cs}$ \citep{schaye2001}. Here $G$ is the gravitational constant, $\rho$ is the matter density, $L$ is the characteristic size of the cloud, and $c_{\rm s}$ ($\equiv \sqrt{{\rm d}P/{\rm d}\rho}$, where $P$ is the pressure) is the speed of sound within the medium.\footnote{This is the same argument as that applied to the Jeans length.} Accounting for helium abundance, this characteristic size is \citep{schaye2001}:
\begin{align}
    L \approx \frac{c_{\rm s}}{\sqrt{\rho G}} 
              \sim 0.52 \ {\rm kpc} \ 
              \left(\frac{ n_{\rm H}}{1\,{\rm cm}^{-3}}\right)^{-1/2}\left( \frac{T}{10^4\,{\rm K}} \right)^{1/2} 
              \left(\frac{f_{\rm g}}{0.17} \right)^{1/2} \, ,
\end{align}

\noindent 
where $f_{\rm g}$ is the gas mass fraction. In the context of the IGM, dark matter dominates the matter content and the gas fraction should be $f_{\rm g}\sim \Omega_{\rm b}/\Omega_{\rm m}\approx 0.17$ \citep{planck2020}, 
while the typical IGM temperature is $T=10^4$\,K. Similarly, one can consider a characteristic column density as \citep{schaye2001}:
\begin{align}
    \label{eq:Ntot_jeans}
     N_{\rm H} \equiv  n_{\rm H} L_{\rm } \sim 1.6 \times 10^{21} \, {\rm cm}^{-2} \,   \left(\frac{ n_{\rm H}}{1\,{\rm cm}^{-3}}\right)^{1/2}\left( \frac{T}{10^4\,{\rm K}} \right)^{1/2} 
     \left(\frac{f_{\rm g}}{0.17} \right)^{1/2} \, .
\end{align}

\noindent For a highly ionized optically thin medium, given the results from Equations~\ref{eq:case_A}~and~\ref{eq:neutral_frac} and accounting for helium \citep{schaye2001},
\begin{align}
\label{eq:NHI_jeans1}
    N_{\rm HI} &\approx 2.3 \times 10^{13} \, {\rm cm}^{-2}   \left(\frac{n_{\rm H}}{10^{-5}\, {\rm cm}^{-3}} \right)^{3/2} \left(\frac{T}{10^4\,{\rm K}} \right)^{-0.26} \left( \frac{\Gamma_{\rm HI}}{10^{-12}\ {\rm s}^{-1}} \right)^{-1} \left(\frac{f_{\rm g}}{0.17} \right)^{1/2} \, , \\
     N_{\rm HI} & \approx  2.7 \times 10^{13} \, {\rm cm}^{-2} \ (1+\delta)^{3/2} \left(\frac{\Omega_{\rm b}}{0.05}\right)^{3/2}
     \left(\frac{1+z}{4}\right)^{9/2}
     \left(\frac{T}{10^4\,{\rm K}} \right)^{-0.26} \left( \frac{\Gamma_{\rm HI}}{10^{-12}\ {\rm s}^{-1}} \right)^{-1} \left(\frac{f_{\rm g}}{0.17} \right)^{1/2} \, ,
     \label{eq:NHI_jeans2}
\end{align}

\noindent where the densities have been normalized at $z=3$ (see Equation~\ref{eq:evol1}) and the latter expression is written explicitly in terms of the overdensity ($1+\delta$; see Equation~\ref{eq:delta_mult}). This implies that self-gravitating clouds in hydrostatic equilibrium in the IGM at mean cosmological densities and at $T=10^4$\,K would produce absorption lines with neutral hydrogen column densities of $N_{\rm HI} \sim 3 \times 10^{13}$\,\cm\ at $z=3$. Increasing/decreasing the temperature has a limited effect given the mild $T^{-0.26}$ dependence. Similarly, increasing/decreasing either the gas fractions or the photoionization rate, also have limited effects given their $1/2$ and $-1$ exponents. The major dependence's are for redshift evolution and density contrasts. Rewriting Equation~\ref{eq:NHI_jeans2} to solve for overdensity we have,
\begin{align}
\label{eq:density_contrast}
(1+\delta) \approx  \left(\frac{N_{\rm HI}}{10^{13.5}\,{\rm cm}^{-2}}\right)^{2/3}
\left(\frac{\Gamma_{\rm HI}}{10^{-12}\ {\rm s}^{-1}}\right)^{2/3}
\left(\frac{1+z}{4}\right)^{-3}
\left(\frac{T}{10^4\,{\rm K}}\right)^{0.17}
\left(\frac{f_g}{0.17}\right)^{-1/3}
\left(\frac{\Omega_{\rm b}}{0.05}\right)^{-1} \, ,
\end{align}

Equations~\ref{eq:NHI_jeans1},~\ref{eq:NHI_jeans2}~and~\ref{eq:density_contrast} are only valid for the diffuse IGM in the mildly overdense cosmic web, at $0 < \delta  \lesssim 10$ and at redshifts after reionization at $z<6$, where the previous assumptions apply. For instance, if we consider a $1 + \delta \approx 10$ at $z=3$, we expect $N_{\rm HI}\approx 10^{15}$\,\cm\ (fixing all the rest of the parameters), which is no longer an optically thin cloud (see Section~\ref{sec:qal_model_deeper}); more complicated ionization corrections are needed that require radiative transfer solvers \citep[e.g.][]{ferland1998}. However, at $z\approx 0$, an overdensity of $1 + \delta \approx 10$, would correspond to an (almost) optically thin cloud with $N_{\rm HI}\approx 10^{13.5}$\,\cm\ (even considering $\Gamma_{\rm HI}\sim 10^{-13}$\,s$^{-1}$). This illustrates the importance of taking the Universe evolution into account when comparing IGM systems of a given \nhi\ and relating them to overdensities. Finally, at sufficiently large column densities $N_{\rm HI} \gtrsim 10^{19}$\,\cm, self-shielding will completely protect the inner part of the cloud from the external ionizing UV photons, implying a mostly neutral medium ($n_{\rm HI} \sim n_{\rm H}$) and thus, no ionization corrections are needed. In this case, Equation~\ref{eq:Ntot_jeans} would be a more appropriate approximation, i.e. $N_{\rm HI} \approx N_{\rm H}$. \sk

These equations should be considered only as guidelines, and some departures (e.g. the exact normalization and scatter around the relations) are indeed expected. For modeling the WHIM, on the other hand, hydrodynamical simulations are usually required due to the presence of more complex physics, like shocks and instabilities; for instance, \citet{dave2010} find that at a fixed $N_{\rm HI}$, WHIM gas is at a $\approx 5-10 \times$ higher overdensities compared to the diffuse IGM, significantly departing from simply applying Equation~\ref{eq:density_contrast}.


\section{Probing the IGM with quasars}
\label{sec:obs}

\subsection{The physics of absorption lines}\sk
\label{sec:qal_model}

In this Section we provide an overview of the basic physical principles of absorption lines and how we model them.\sk

\subsubsection{Emission and absorption lines}\smallskip

When electromagnetic radiation interacts with ions, there will be physical processes that can produce what we call emission and absorption lines. These are produced at very specific energies, given by quantum mechanics, which translates to specific frequencies or wavelengths in an spectrum. The physical systems that produce emission or absorption lines are usually referred to as `emitters' and `absorbers', respectively. These can be individual ions, or, more typically, ensembles of them (e.g. ``clouds'' of gas or plasma).\sk

Let us consider a cloud of ions $X$ that has two quantized energy levels: an upper level with energy $E_u$, and a lower level with energy $E_l$. If photons of frequencies $\nu_{ul}=(E_u-E_l)/h$ are present, then these can be absorbed by the ions in the lower energy level, $X_l$, and produce ions excited to the upper level, $X_u$:
\begin{align}
\label{eq:absorption}
    X_l + h\nu_{ul} \rightarrow X_u \, .
\end{align}

\noindent This process is called ``absorption'', and the rate at which this happens is proportional to the number density of $X_l$ ions, $n_l$, and the specific energy density, $u_{\nu}$:
\begin{align}
\label{eq:abs_rate1}
    \left(\frac{{\rm d}n_{u}}{\rm{d}t}\right)_{l \rightarrow u} =  - \left(\frac{{\rm d}n_{l}}{\rm{d}t}\right)_{l \rightarrow u} =  n_l B_{lu} u_{\nu} \, ,
\end{align}
\noindent where the constant of proportionality, $B_{lu}$ is the Einstein $B$ coefficient for this particular transition from $X_l$ to $X_u$. \sk

In contrast to absorption, ``emission'' is produced when the $X_u$ ions gets de-excited from the upper to the lower level, by emitting a photon with frequencies $\nu_{ul}=(E_u-E_l)/h$. There are two ways in with process can happen: spontaneous emission and stimulated emission,
\begin{align}
    X_u & \rightarrow X_l + h\nu_{ul} \, ,\\
    X_u + h\nu_{ul} & \rightarrow  X_l +2h\nu_{ul} \, ,
\end{align}
\noindent respectively. As its name suggests, spontaneous emission is independent of the incident radiation field, and its rate is simply proportional to the number density of $X_u$ ions, $n_u$. On the other hand, stimulated emission does depend on the specific energy density of the incident radiation, $u_{\nu}$. Thus, combining these two contributions, the rate at which emission from $X_u$ to $X_l$ happens, is:
\begin{align}
    \left(\frac{{\rm d}n_{l}}{\rm{d}t}\right)_{u \rightarrow l} =  - \left(\frac{{\rm d}n_{u}}{\rm{d}t}\right)_{u \rightarrow l} =n_u (A_{ul} +B_{ul} u_{\nu}) \, ,
\end{align}

\noindent where the constants of proportionality, $A_{ul}$ and $B_{ul}$, are the Einstein's $A$ and $B$ coefficient for this particular transition (from $X_u$ to $X_l$). Assuming thermal equilibrium between the radiation and the ions, the ratio $n_u/n_l$ is given by the Boltzmann equation $n_u/n_l=(g_u/g_l)e^{-(E_u-E_l)/kT}$, where $g_u$ and $g_l$ are the statistical weights for the upper and lower level, respectively. Then, it can be shown that the three Einstein's coefficients are related to each other as \citep{draine2011}
\begin{align}
   B_{ul} &= \frac{c^3}{8\pi h\nu_{ul}^3}A_{ul}  \\
    B_{lu} &= \frac{g_u}{g_l}B_{ul} \, ,
\end{align}

Stimulated emission can be ignored in the IGM conditions, and thus we only need to consider spontaneous emission. In this case, the rates can be expressed as:
\begin{align}
\label{eq:abs_rate2}
\left(\frac{{\rm d}n_{u}}{\rm{d}t}\right)_{l \rightarrow u} &= n_l \frac{g_u}{g_l}\frac{c^3}{8\pi h\nu_{ul}^3}A_{ul} \, , \\
\left(\frac{{\rm d}n_{l}}{\rm{d}t}\right)_{u \rightarrow l} &= n_uA_{ul}  \, .
\label{eq:em_rate2}
\end{align}

For specific ions and transitions, the Einstein's $A$ and $B$ coefficients can be determined from quantum mechanics and/or directly from experiments, and are usually tabulated (see \citealt{draine2011} and references therein).

\subsubsection{Modeling absorption lines}\smallskip
\label{sec:qal_model_deeper}

\paragraph{Absorption cross-section and the line profile}\sk
Interpreting absorption lines (e.g. those in the spectra of a background source that provides a continuum) requires modeling them in terms of the observed line profile and the absorption cross-section. In this case, the absorption cross-section, $\sigma_{lu}(\nu)$, can be defined by re-writing Equation~\ref{eq:abs_rate1} as (similar to the rates defined in Section~\ref{sec:ionization}),
\begin{align}
\label{eq:abs_rate3}
    \left(\frac{{\rm d}n_{u}}{\rm{d}t}\right)_{l \rightarrow u} =  n_l \int \frac{u_{\nu}}{h\nu} \sigma_{lu}(\nu) c \, {\rm d}\nu  \approx  n_l  \frac{u_{\nu}}{h\nu_{ul}} c\int \sigma_{lu}(\nu) \,  {\rm d}\nu \, ,
\end{align}
\noindent where the latter approximation assumes that $u_{\nu}$ and $h\nu$ are approximately constants over the absorption line profile given by $\sigma_{lu}(\nu)$. Comparing Equation~\ref{eq:abs_rate3} to Equations~\ref{eq:abs_rate1}~and~\ref{eq:abs_rate2}, we can relate the absorption cross-section to the Einstein's coefficients, as
\begin{align}
    \int \sigma_{lu}(\nu)\, {\rm d}\nu = \frac{h\nu_{ul}}{c}B_{lu} = \frac{g_u}{g_l}\frac{c^2}{8\pi \nu_{ul}^2}A_{ul} \, .
\end{align}

\noindent From this, we can have an expression for the absorption cross-section in terms of the ``natural line profile'', $\phi_{\rm nat.}(\nu)$, as
\begin{align}
\label{eq:sigma_cross_1}
     \sigma_{lu}(\nu) = \frac{g_u}{g_l}\frac{c^2}{8\pi \nu_{ul}^2}A_{ul}\phi_{\rm nat.}(\nu)  \, ,
\end{align}
\noindent where $\phi_{\rm nat.}(\nu)$ is a normalized function that encodes only the shape of the spectral line, i.e. it satisfies $\int \phi_{\rm nat.}(\nu)\, {\rm d}\nu=1$. \sk

At this point it is important to note that despite what its name may imply, an ``absorption line'' can be either produced by actual absorption of the photons into individual ions (excitation or ionization) and/or scattering (effectively an absorption and instantaneous re-emission), i.e. the ions just change the original trajectory of photons and scatter them away from the line-of-sight. In both cases the net effect is a deficit of photons from the background spectrum at specific frequencies producing an absorption line.

\paragraph{Oscillator strength}\sk
Another useful manner to quantify the strength of an absorption line produced by an ion in $X_l$ being excited to $X_u$ (Equation~\ref{eq:absorption}) is via the ``oscillator strength'', $f_{lu}$. The oscillator strength is dimensionless and it corresponds to the probability of a transition between two quantum states due to interaction with electromagnetic radiation. In terms of the absorption cross-section it can be defined as,
\begin{align}
    f_{lu}  = \frac{m_ec}{\pi e^2}  \int \sigma_{lu}(\nu)\, {\rm d}\nu \, ,
\end{align}
\noindent and thus it is related to the natural absorption line profiles as:
\begin{align}
\label{eq:fosc_def2}
     \sigma_{lu}(\nu) &= \frac{\pi e^2}{m_e c} f_{lu} \phi_{\rm nat.}(\nu) \, .
\end{align}
\noindent Of course, one can also find expressions of $f_{lu}$ in terms of Einstein's coefficients (not done here, but see \citealt{draine2011}). For emission lines, the oscillator strength is defined to be negative, and the corresponding value for the particular emission line satisfies $f_{ul} = - \frac{g_l}{g_u} f_{lu}$. Similarly to the Einstein coefficients, the values of the oscillator strengths for different ion transitions can be computed from quantum mechanics and/or measured from experiments \citep{verner1994,morton2003, draine2011}. 

\paragraph{Line broadening}\sk
In practice, the {\it observed} line profile may differ from that of the natural profile, because of different physical processes. In the following we summarize the most important types of line broadening concerning IGM observations:\footnote{But note that in other astrophysical contexts, other types of line broadening  can be important too (e.g. collisional broadening).} \sk
\begin{itemize}
    \item \textbf{Natural broadening:} The shape of $\phi_{\rm nat.}(\nu)$ is driven by quantum mechanics. Although the line will be centred at frequency $\nu_{ul}$, because of the uncertainty principle, $\Delta E \Delta t \ge \frac{h}{4\pi}$, transitions from energy levels with larger lifetimes will have narrower line profiles, and vice-versa. The intrinsic natural line profile is well characterized by a Lorentzian function with a full-width at half maximum (FWHM) determined by the Einstein's $A$ coefficients:
\begin{align}
    {\rm FWHM}_{\phi_{\rm nat.}} = \frac{1}{2\pi}\left(\sum_{E_i<E_u} A_{ui} + \sum_{E_i<E_l} A_{li} \right) \, .
\end{align}
\noindent Note that for transitions where $l=0$ (ground state) the second terms is zero. The Lorentzian profile is characterized by the damping parameter, $\gamma_{lu}$ ($=\gamma_{ul}$), which is related to the FWHM as ${\rm FWHM}_{\phi_{\rm nat.}} = 2\pi/\gamma_{lu}$.\sk
    
    \item \textbf{Thermal broadening:}  Thermal broadening (also known as Doppler broadening) occurs due to the random thermal motion of ions within the absorber. Each ion's rest-frame frequency of absorption (or emission) shifts to a different observed frequency because of the Doppler effect, which leads to a broadening of the spectral line. Assuming a Maxwellian velocity distribution for the ions, the resulting thermal line profile is Gaussian, with a a corresponding $1$-dimensional velocity standard deviation, $\sigma_v=\sqrt{\frac{kT}{m}}$, where $m$ is the mass of the ion. The FWHM of the line profile is ${\rm FWHM}_{\phi_{\rm th}}=2\sqrt{\ln{2}}b_{\rm th}$, with $b_{\rm th}$ being the so-called ``Doppler parameter'' defined as $b_{\rm th} \equiv \sqrt{2}\sigma_{v}$.\sk
    
    \item \textbf{Turbulent broadening:} Turbulent broadening is due to macroscopic motions within the absorber. While the exact nature and modeling of turbulence in astrophysical environments remain areas of active research, turbulent broadening is usually characterized by a ``turbulent Doppler parameter'', $b_{\rm turb.}$, which is treated similarly to that of the thermal broadening such that the total Doppler parameter can be expressed as $b=\sqrt{b^2_{\rm th} + b^2_{\rm turb.}}$. However, in contrast to the thermal broadening, turbulent broadening is independent of the atomic mass of the ions, since it reflects the bulk motion of the gas as a whole rather than the random thermal motions of the individual particles.\sk
    \item \textbf{Instrumental broadening:} Instrumental broadening (also known as the `line spread Function', LSF) results from the finite resolution of spectrographs due to the combined telescope and instrument optics plus image quality (including the effects of the atmosphere, if any, e.g. the `seeing'). The instrumental profile sets the narrowest possible spectral line that can be observed, meaning that lines with physical FWHM smaller than this limit are {\it unresolved}. Typically, the instrumental broadening profile is characterized and known for a given spectrograph and thus can be properly accounted for when modeling absorption or emission lines. The shape of the LSF can vary from simple Gaussian profiles (most common case), to more complex shapes with no analytical description and thus are given as tabulated values instead.\footnote{e.g. The Cosmic Origin Spectrograph LSF is described in \url{ https://www.stsci.edu/hst/instrumentation/cos/performance/spectral-resolution}}
\end{itemize}


\paragraph{The Voigt profile}\sk
Given that the natural, thermal, and turbulent broadening processes are all present when observing absorption lines, the {\it actual} line profile ends up being a combination of Gaussian and Lorentzian profiles. This combination results in what is known as the Voigt profile, which represents the convolution of these two profiles \citep{vandehulst1947}. Here, we will refer to this absorption line profile simply as $\phi_{\nu}^{\rm Voigt}$, which also satisfies $\int \phi_{\nu}^{\rm Voigt}\, {\rm d}\nu=1$. In general, the Voigt profile should be used when measuring absorption lines for determining the properties of the absorbers. For this, dedicated open-source software is available for general use \citep[e.g.][]{vpfit, hummels2017}. Finally, the information from an individual Voigt profile fit is usually stored as a $(z_{\rm HI}, N_{\rm HI}, b_{\rm HI})$ tuple.\sk

\begin{figure}[t]
\centering
\includegraphics[width=.7\textwidth]{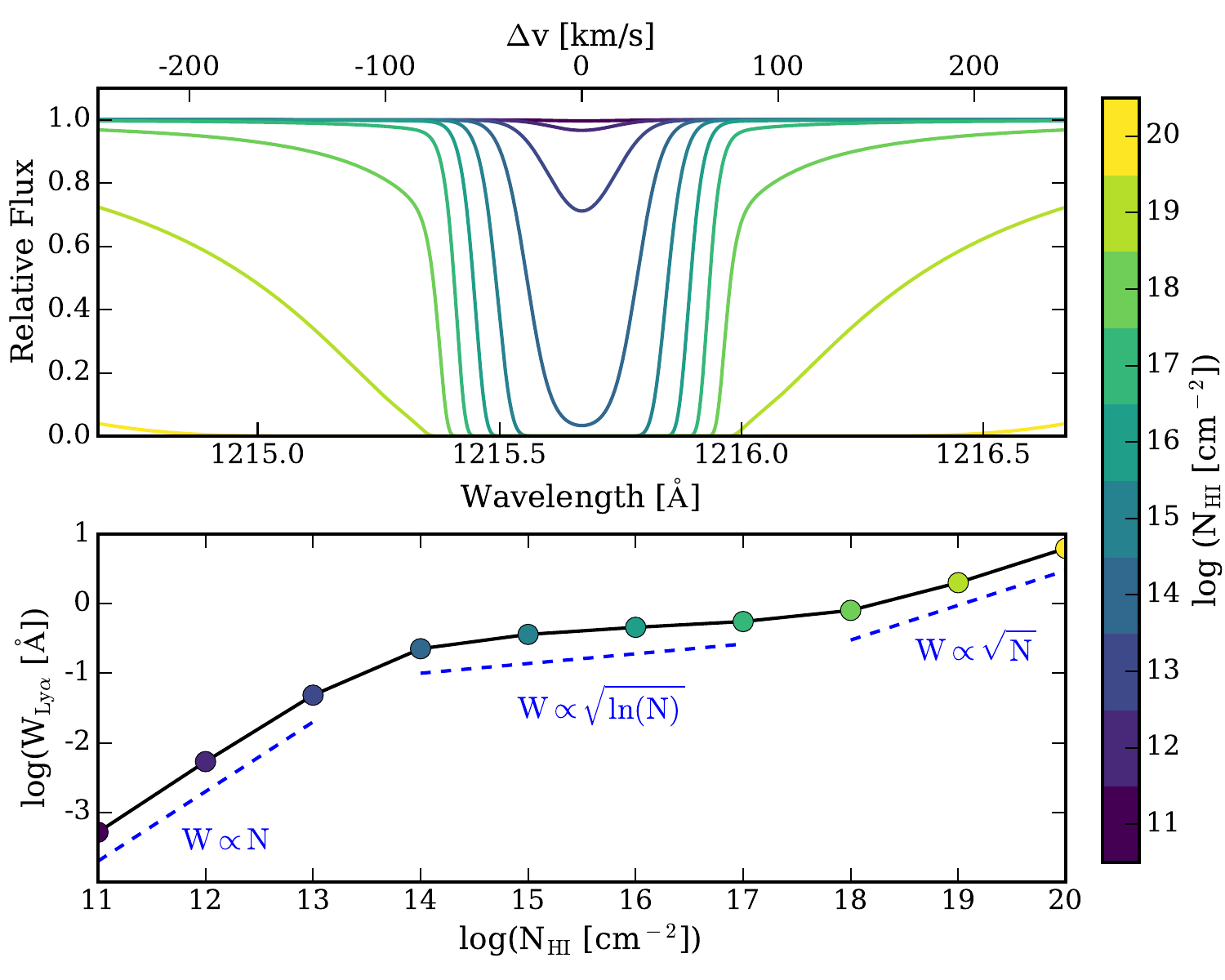}
\caption{Examples of theoretical absorption line profiles for the \hi~\lya\ transition in a normalized spectrum (top) as a function of column density (see color bar) fixing the Doppler parameter to $b=22$\kms, and the corresponding Curve of Growth (bottom). Figure from \citet{hummels2017}.}
\label{fig:trident}
\end{figure}

Examples of Voigt profiles for \hi~\lya\ with different column densities are given in Figure~\ref{fig:trident} (top panel). Weak (optically thin) absorption lines are dominated by the Gaussian component (in the core), while extremely saturated lines are dominated by the Lorentzian component instead (in the damping wings). In these extreme cases, the properties of the absorbers can often be directly inferred even from individual absorption lines (see `Curve of Growth', below). For intermediate cases, there is a degeneracy between the intrinsic parameters, and thus more than one transition is usually required to determine column densities accurately. For more details about this Voigt profile, we refer the reader to \citet{draine2011}.

\paragraph{Optical depth, column density, and equivalent width}\sk

In order to link the previous modeling of absorption lines with actual observations, we need to consider the radiative transfer equation. When the light of a quasar (or any other background source) passes through an intervening absorber, then its continuum intensity, $I_{\nu}$, will be attenuated in proportion to both the absorption cross-section ($\sigma_{lu}$) and the density of ions that can produce the absorption ($n_l$). In the IGM conditions, the radiative transfer equation is simply (ignoring the emissivity term),
\begin{align}
\label{eq:transfer}
    {\rm d}I_{\nu} = -I_{\nu}\kappa_{\nu}\,{\rm d}s \, ,
\end{align}

\noindent where $\kappa_{\nu} \equiv n_{l}\sigma_{lu}(\nu)$ is the attenuation coefficient at frequency $\nu$, and ${\rm d}s$ is the infinitesimal pathlength of propagation. For integrating this equation, it is convenient to define the ``optical depth'', $\tau_{\nu}$, as satisfying,
\begin{align}
\label{eq:tau_def}
    {\rm d}\tau_{\nu} \equiv \kappa_{\nu}\,{\rm d}s \, ,
\end{align}

\noindent such that Equation~\ref{eq:transfer} becomes,
\begin{align}
\label{eq:transfer2}
    {\rm d}I_{\nu} = -I_{\nu}{\rm d}\tau_{\nu} \, .
\end{align}

\noindent The solution is then,
\begin{align}
\label{eq:X}
    I_{\nu}(\tau_{\nu}) = I_{\nu}(0)\exp{(-\tau_{\nu})} \, ,
\end{align}
\noindent were the $I_{\nu}(0)$ represents the intensity of radiation in the absence of the absorber. Assuming that the absorber completely covers the beam of the quasar, then we have a similar equation for the flux density, $F_{\nu}$,
\begin{align}
\label{eq:X}
    F_{\nu}(\tau_{\nu}) = F_{\nu}(0)\exp{(-\tau_{\nu})} \, ,
\end{align}

\noindent where $F_{\nu}(0)$ is the flux density of the source in the absence of the absorber, i.e. $F_{\nu}(\tau_{\nu}=0)$. Thus, the quasar continuum spectrum will be exponentially attenuated at very specific frequencies (or wavelengths)  given by the optical depth,
\begin{align}
\label{eq:X}
    \tau_{\nu} &= \int \kappa_{\nu}\, {\rm d}s \\ 
\label{eq:tau_def_2}
    & = \int n_l \sigma_{lu}(\nu) \, {\rm d}s \\ 
    & = \frac{\pi e^2}{m_e c} f_{lu} \int n_l\phi^{\rm Voigt}_{\rm \nu}\, {\rm d}s\, ,
\label{eq:tau_def_3}
\end{align}

\noindent where we have used Equations~\ref{eq:tau_def}~and~\ref{eq:fosc_def2}, respectively. \sk

Given that the absorber can have a particular density profile, the density $n_l$ should be a function of the pathlength, i.e. $n_l(s)$. However, the Voigt profile only depends on quantum mechanical properties of the ions, and the macroscopic properties of the absorber, like temperature and turbulence. It is usually assumed that within a single resolved absorption line, the temperature and turbulence are roughly constant, such that, $\phi_{\nu}^{\rm Voigt}$ is roughly independent of the pathlength, and thus
\begin{align}
\label{eq:X}
    \tau_{\nu} &\approx \frac{\pi e^2}{m_e c} f_{lu} \phi^{\rm Voigt}_{\rm \nu} \int n_l\,{\rm d}s \\
    &= \frac{\pi e^2}{m_e c} f_{lu} \phi^{\rm Voigt}_{\rm \nu} N_{l} \, ,
\end{align}

\noindent where,
\begin{align}
    N_l \equiv  \int n_l{\rm d}s \, ,
\end{align}

\noindent is the column density. 
Therefore, the optical depth is directly proportional to the column density, the oscillator strength, and the Voigt profile,
\begin{align}
    \tau_{\nu} \propto N_l f_{lu} \phi^{\rm Voigt}_{\rm \nu} \, .
\end{align} 
\noindent Similarly, if the density does not vary significantly along the pathlength, we can further infer that,
\begin{align}
\label{eq:tau_nu_constant}
    \tau_{\nu} \propto n_l f_{lu} \phi^{\rm Voigt}_{\rm \nu} \, \, ({\rm for} \, n_l = {\rm const.}).
\end{align}

Assuming the absorption lines are fully resolved by the spectrograph (i.e. the LSF is much narrower than the other line broadening mechanisms), then the Voigt profile is mostly determined by the quantum mechanical properties of the ion transition and the thermal (and/or turbulent) state of the absorber. In this case, we can use the shape and intensity of the absorption line to infer important physical parameters like their column densities and temperatures (and/or turbulence). However, in practice, because we do not always fully resolve the absorption profiles and/or the signal-to-noise ($S/N$) of the spectra is too low for properly characterizing the shape of the absorption line, these physical quantities are not readily accessible from this treatment. For these cases (and other uses) it is convenient to define the ``equivalent width'', $W$, as
\begin{align}
    W &\equiv  \int \frac{F_{\lambda}(0)-F_{\lambda}(\tau_{\lambda})}{F_{\lambda}(0)} \, {\rm d}\lambda \\
    & = \int (1-\exp{(-\tau_{\lambda})}) \, {\rm d}\lambda \, ,
\end{align}
\noindent where $\tau_{\lambda}$ is the optical depth expressed as a function of wavelength, as we typically work with spectra given as a function of wavelength, $F_{\lambda}$. The equivalent width corresponds to the amount of flux ``absorbed'' relative to the continuum and has units of length.\footnote{For emission lines, the equivalent width is (by convention) negative, and it represents the amount of flux emitted relative to the continuum.} It is easy to show that $W$ is independent of the resolving power of the spectrograph, which makes it a versatile observable which can (in some cases) be linked back to the more physical quantities of interest (e.g. column density) based on the ``Curve of Growth''.

\paragraph{Curve of Growth}\sk

The relation between equivalent width ($W$) and column density ($N_l$) for a given ion transition is known as the Curve of Growth (COG). Given the definition of $W$, and the fact that $\tau_{\lambda} \propto N_{l}\phi^{\rm Voigt}_{\rm \lambda}$, then passing from known values of $N_l$ and $b$ to $W$ is unambiguous. However, in practice one usually measures $W$ and wishes to infer $N_l$, and the relation will depend on the value of $b$. Figure~\ref{fig:trident} (bottom panel) shows an example of a COG for the \hi~\lya\ transition for a fixed Doppler parameter $b$. The optical depth at line-center, $\tau_0$, can be approximated by \citep{draine2011}
\begin{align}
\label{eq:tau0}
    \tau_0 & \approx \sqrt{\pi}\frac{e^2}{m_e c}\frac{N_l f_{lu}\lambda_{lu}}{b} \, ,
\end{align}

\noindent where $\lambda_{lu}$ is the wavelength associated to the transition. Here, it is important to emphasize that $\tau_0 \propto N_{l}/b$, thus for a given ion with a given $\tau_0$, broader absorption lines trace higher column densities.\sk 

Based on the value of $\tau_0$, the COG can be divided into three main regimes, each having a different dependence between $W$ and $N_{l}$ as follows:\sk

\begin{itemize}
    \item {\bf Linear regime:} For $\tau_0 \lesssim 1$ (optically thin), the COG has a linear $ W \propto N_{l}$ dependence, which is actually independent of $b$.\sk
    
    \item {\bf Flat regime:}  When $10 \lesssim \tau_0  \lesssim 10^3$ (optically thick), the COG has a $W \propto b \sqrt{\ln{(N_{l}/b)}}$, hence a very mild (almost flat) dependence of $W$ as a function of $N_{l}$.  In this regime, the COG depends on $b$, and thus $W$ is degenerate for some combinations of $N_{l}$ and $b$.\sk

    \item {\bf Square root regime:}  When $10^4 \lesssim \tau_0$ (optically thick), the COG has a $W \propto \sqrt{N_{l}}$ dependence, which is (again) independent of $b$. \sk
    
\end{itemize}

\noindent Therefore, for the linear and square root regimes, it is relatively straightforward to infer $N_l$ from $W$, even for isolated unresolved lines. However, this is generally not possible to do in the flat regime; even if one knows the value of $b$ precisely, a small error in $W$ propagates into a large error in $N_l$. For this latter case, one should attempt to use another transition(s) of the same ion falling into either the linear or the square root regime.\footnote{For instance, for \hi, if \lya\ is in the flat regime, one could use \lyb\ or higher order Lyman series instead.}\sk

Although the exact shape of the COG requires a proper integration of the Voigt profile (non-trivial), \citet{draine2011} provides a useful analytical approximation as
\begin{align}
\label{eq:draine_cog}
W \approx
\begin{cases} 
   \, \lambda_{lu}\sqrt{\pi} \frac{b}{c}\frac{\tau_0}{1+\tau_0/(2\sqrt{2})}  & \text{if } \tau_0 \le 1.25393 \, , \\
   \,  \lambda_{lu} \left[\left(\frac{2b}{c}\right)^2\ln{\left(\frac{\tau_0}{\ln{2}}\right)} + \frac{b}{c}\frac{\gamma_{lu}\lambda_{lu}}{c}\frac{(\tau_0-1.25393)}{\sqrt{\pi}}  \right]^{1/2}& \text{if } \tau_0 > 1.25393 \, .
\end{cases}
\end{align}

\noindent For instance, evaluating Equation~\ref{eq:tau0} for the \hi~\lya\ transition at typical column density of $N_{\rm HI} = 10^{13.5}$\,\cm\ and typical Doppler parameter of $b=20$\,\kms, we obtain,
\begin{align}
\label{eq:tau0_eval}
     \tau_0  = 1.255 \left(\frac{N_l}{10^{13.5}\, {\rm cm}^{-2}}\right)\left(\frac{f_{lu}}{0.4164}\right)\left(\frac{\lambda_{lu}}{1215.67\, \AA}\right)\left(\frac{b}{20\,{\rm km\,s}^{-1}}\right)^{-1} \, ,
\end{align}
\noindent coinciding with the approximate limit between the linear and the flat portions of the COG (Equation~\ref{eq:draine_cog}). In contrast, the limit between the flat and square root portion of the COG will be at a $\approx 10^4$ larger column density, i.e. at $N_{\rm HI} = 10^{17.5}$\,\cm\ (for the same $b=20$\,\kms). These (and other) useful approximations and relations for the modeling and treatment of absorption lines are included in the open-source Python package \linetools\ \citep{linetools} for general use.

\subsection{Observations of the IGM through quasar absorption lines}\sk
\label{sec:qal}

Quasar absorption lines have been extensively used to probe and study the IGM. Quasars allow us to probe the Universe since the EoR until the present day \citep{fan2023}. In this Section, we will present a non-exhaustive review of strategies and observational methods for probing the IGM with quasar absorption lines.\sk

\subsubsection{The Gunn-Peterson effect}\sk
\label{sec:gunn-peterson}

As mentioned in Section~\ref{sec:history}, an important early discovery was that of \citet{gunn1965}: the fact that we observe transmitted flux blueward of the \lya\ emission of high-$z$ quasars imply an extremely highly ionized IGM. The argument is simple but very powerful, summarized as follows.\sk 

The optical depth ($\tau_{\nu}$) of a single \hi~\lya\ can be written as (Equation~\ref{eq:tau_def_2}),
\begin{align}
    \tau_{\nu} & = \int_0^s n_{l}(s) \sigma_{lu}(\nu) \, {\rm d}s \\ 
               & =  \sigma_{{\rm Ly}\alpha} \nu_{ul}\int_0^s n_l(s)\phi^{\rm Voigt }(\nu)  \, {\rm d}s \, ,
\end{align}
\noindent where $\sigma_{{\rm Lya}}\equiv \frac{\pi e^2}{m_e c} \frac{f_{lu}}{\nu_{ul}} 
= 4.48\times 10^{-18}$\,cm$^2$ is the \hi~\lya\ cross-section. If we consider the absorber to be at a cosmological redshift $z$, then $\frac{{\rm d}s}{{\rm d}z} = \frac{c}{H(z)(1+z)}$, and thus,
\begin{align}
    \tau_{\nu} & =  c \sigma_{{\rm Ly}\alpha} \nu_{ul}\int_0^z n_l(z)\phi^{\rm Voigt }(\nu)  \, \frac{{\rm d}z}{H(z)(1+z)} \, , 
\end{align}

\noindent with $H(z) = H_0\sqrt{\Omega_{\rm m}(1+z)^3 + \Omega_{\Lambda}}$ for a flat \lcdm\ cosmological model. Given that the line profile, $\phi^{\rm Voigt}(\nu)$, strongly peaks at $\nu=\nu_{ul}/(1+z)$, then,
\begin{align}
    \tau_{{\rm Ly}\alpha}  & =  \sigma_{{\rm Ly}\alpha} \frac{c}{H_0}\frac{n_{\rm HI}(z)}{\sqrt{\Omega_{\rm m}(1+z)^3 + \Omega_{\Lambda}}} \, \\
      & = \sigma_{{\rm Ly}\alpha} \frac{c}{H_0}\frac{\langle n_{\rm H} \rangle_0(1+z)^3}{\sqrt{\Omega_{\rm m}(1+z)^3 + \Omega_{\Lambda}}} x_{\rm HI}(z) \, ,
\end{align} 
\noindent where we have renamed $n_{l}$ as $n_{\rm HI}$, and $\tau_{\nu}$ as $\tau_{{\rm Ly}\alpha}$, and $x_{\rm HI}(z)$ is the corresponding neutral fraction. Evaluating this equation for $H_0=70\, {\rm km\,s}^{-1}{\rm \,Mpc}^{-1}$ and $\langle n_{\rm H} \rangle_0=2.5 \times 10^{-7}$\,\cc\ (Equation~\ref{eq:n0}), we obtain, 
\begin{align}
 \tau_{{\rm Ly}\alpha}(z) = 1.48 \times 10^{4} \, x_{\rm HI}(z) \, \frac{(1+z)^3}{\sqrt{\Omega_{\rm m}(1+z)^3 + \Omega_{\Lambda}}} \, ,
\end{align}
\noindent or,
\begin{align}
    x_{\rm HI}(z) = 6.8 \times 10^{-5} \, \tau_{{\rm Ly}\alpha}(z)  \frac{\sqrt{\Omega_{\rm m}(1+z)^3 + \Omega_{\Lambda}}}{(1+z)^3} \, .
\end{align}

\noindent Observations of the \lya~forest at $z\gtrsim 3$ indicate $\tau_{{\rm Ly}\alpha} \lesssim 1$, while at $z\lesssim 1$ the $\tau_{\rm Lya}$ are even smaller (see Figure~\ref{fig:lyaforest}). This implies that the $x_{\rm HI}$ must be extremely low \citep{gunn1965}: as reference, a $\tau_{{\rm Ly}\alpha}\sim 1$ at $z=\{0, 3, 6\}$, corresponds to $x_{\rm HI}\sim \{7, 0.5,0.2\}\times 10^{-5}$, respectively. This is what is known as the `Gunn-Peterson effect': only $1$ neutral hydrogen per $10^5$ or $10^6$ ionized ones are sufficient to produce significant opacity. At $z\gtrsim 6$, the Universe becomes more neutral, and we can start to see spectral regions that are completely absorbed. These regions are known as `Gunn-Peterson troughs' (e.g. see the highest redshift quasar in Figure~\ref{fig:lyaforest}; see also \citealt{fan2023}), yet these could still be produced by a highly ionized medium, with $x_{\rm HI}\sim 10^{-4}$. Flux transmission spikes between Gunn-Peterson troughs can be used to establish the sizes of early ionizing bubbles and constrain models for the ionizing sources \citep{fan2023}.\sk

\subsubsection{Observational classification of \hi\ absorbers}\sk
\label{sec:qal_class}

\begin{figure}[t]
\centering
\includegraphics[width=1.0\textwidth]{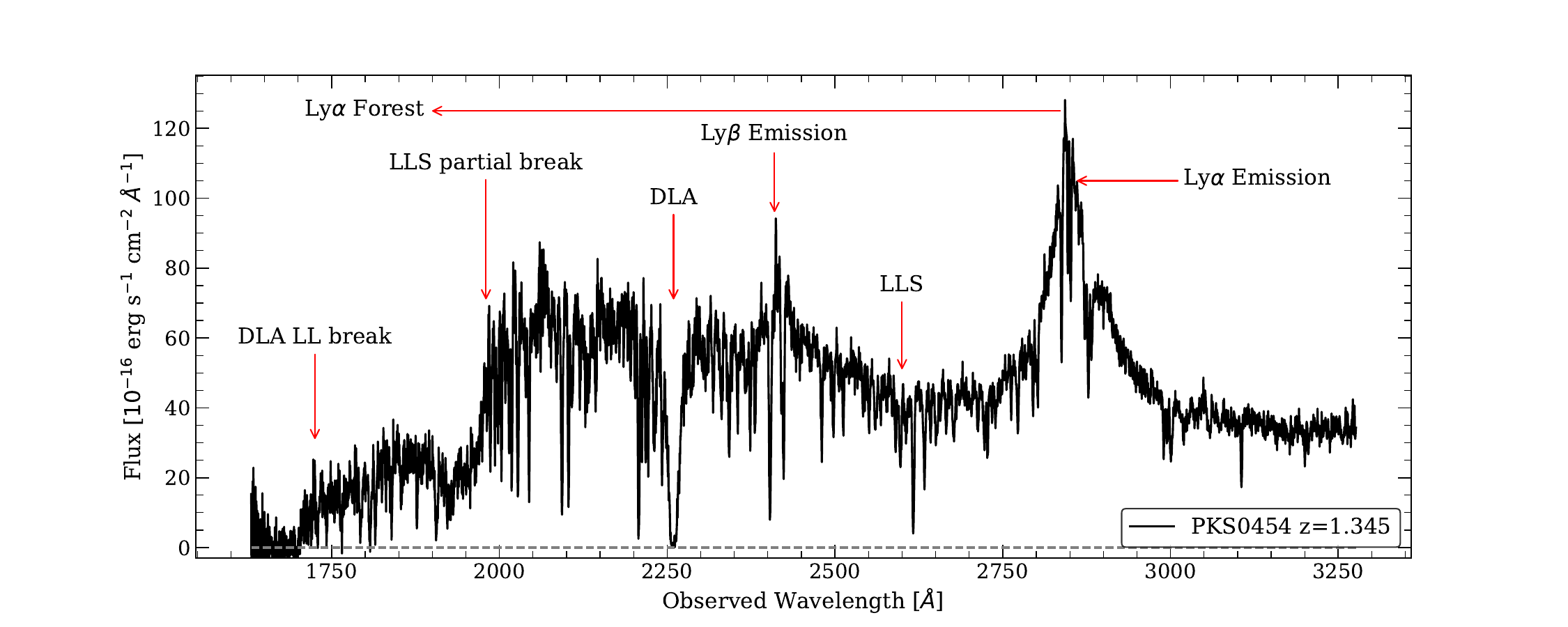}
\caption{Example of a quasar spectrum (PKS\,0454 at $z=1.345$), where the different types of (intervening) \hi\ absorbers are indicated (see Section~\ref{sec:qal_class}). The \hi~\lya\ and \lyb\ emission of the quasar itself are also marked. Figure credit:  K. Mart\'inez-Acosta (adapted from fig. 1 of \citealt{charlton2000}).}
\label{fig:qso_absorbers}
\end{figure}

With sufficient spectral resolution, individual \hi\ absorption lines can be resolved and we can study the different population of absorbers in more detail. Historically, \hi\ absorbers are classified into different categories depending on their observational signatures imprinted in the quasar spectrum (e.g. Figure~\ref{fig:qso_absorbers}). More quantitatively, this classification depends on their column density as follows:\sk

\begin{itemize}
    \item {\bf \lya~forest systems:} correspond to the weakest \hi~\lya\ absorption lines at column densities of $N_{\rm HI} \lesssim 10^{17.2}$\,\cm. These are the most common absorption lines, and trace environments around the mean density of the Universe and up to mild overdensities $(1+\delta) \lesssim 10$ (see Equation~\ref{eq:density_contrast}), i.e. the diffuse IGM. Their name comes from the fact that when observing a quasar spectrum, these absorption lines appear blueward of the quasar's \lya\ emission resembling a crowded ``forest'' (see Figures~\ref{fig:lyaforest}~and~\ref{fig:qso_absorbers}). A review article of these systems was written by \citet{rauch1998}.\sk
    
    \item {\bf Lyman Limit systems (LLSs):} correspond to \hi~\lya\ absorption lines at column densities $10^{17.2} \lesssim N_{\rm HI} \lesssim 10^{20.3}$\,\cm. At these higher column densities, it is possible to appreciate the Lyman limit break, a discontinuity at $912$\,\AA\ (in the rest-frame; equivalent to $13.6$\,eV) in the quasar spectrum, due to the lack of photons that are capable of fully ionizing the \hi\ atoms in the absorber (see Figure~\ref{fig:qso_absorbers}). Depending on whether the break reaches the zero level flux or not, these are referred to as LLSs or partial LLSs (pLLSs), respectively. The column density of $10^{17.2}$\,\cm\ corresponds to that of the inverse of the \hi\ photoionization cross-section of $\sigma_{\rm p.i.}=6.3 \times 10^{-18}$\,cm$^2$ (Equation~\ref{eq:sigma_pi}). In terms of over-densities, these typically trace gas within or nearby galaxy halos, at $1+\delta \gtrsim 100$. Some relevant articles about LLSs include \citet{bahcall1965,tytler1982,songaila2010, prochaska2010,fumagalli2011, mcquinn2011,lehner2013}.\sk

    \item {\bf Damped Lyman-$\alpha$ systems (DLAs):} correspond to \hi~\lya\ absorption lines at column densities $N_{\rm HI} \gtrsim 10^{20.3}$\,\cm. At these high column densities, it is possible to clearly appreciate the damping wings of the Lorentzian profile of the \hi~\lya\ absorption line, even in low-resolution data (see Figure~\ref{fig:qso_absorbers}). This column density limit also corresponds to the transition towards a mostly neutral medium. In terms of over-densities, these are typically associated with environments near galaxy disks or within galaxies themselves (e.g. the ISM). A review article of these systems was written by \citet[][see also \citealt{peroux2020}]{wolfe2005}.\sk
\end{itemize}

\noindent Despite this classification scheme being somewhat arbitrary (based on properties seen in the spectra of quasars), it is nevertheless sufficiently convenient that it is still being widely used. These categories resemble those of the COG (although not exactly). \lya~forest absorbers are associated with the linear and flat portions of the COG, while LLS and DLAs are associated to the square root portion of it. Among these three, the LLSs are the most difficult to model given  the uncertainties in the ionization corrections (among other issues); in contrast, DLAs are mostly neutral absorbers, while \lya~forest systems are mostly ionized. For characterizing LLS, it is usually required to have several of the Lyman series lines modeled simultaneously (covering different regimes in the COG), and also include the Lyman Limit break intensity. Some authors even define an intermediate category between LLSs and DLAs, at $10^{19} \lesssim N_{\rm HI} \lesssim 10^{20.3}$\,\cm, referred to as super-LLS or sub-DLAs systems, which speaks of the complexity of the transition from LLS to DLA \citep{peroux2020}.\sk

\subsubsection{Surveys of \hi\ absorbers}\sk
\label{sec:qal_survey}

In order to study these \hi\ absorbers at different redshifts and/or environments, we require observational surveys. A survey is typically constructed by using a sample of spectra of  background quasars to search and characterize absorption lines. The majority of these absorption lines at observed wavelengths shorter than the \hi~\lya\ emission of the quasar, $\lambda_{\rm obs} < \lambda^{\rm Ly\alpha}_{\rm qso} = 1215.67{\rm \AA} (1+z_{\rm qso})$ will correspond to \hi\ (e.g. the \lya-forest), but some can also be produced by helium, metals or even molecules. Indeed, at $\lambda_{\rm obs}>\lambda^{\rm Ly\alpha}_{\rm qso}$ the majority will be from metal ions. These absorption lines will be produced by absorbers at redshifts $z_{\rm abs} \le z_{\rm qso}$, and thus it is convenient to use them to probe different cosmic epochs.\sk

\paragraph{The incidence of \hi\ absorbers}\sk

A key observable is the incidence of absorption lines per unit redshift, 
\begin{align}
 \frac{{\rm d}\mathcal{N}_{\rm abs}}{{\rm d}z} = \frac{\mathcal{N}_{\rm abs}(W_{\rm r}\ge W_{\rm r, min})}{\Delta z (W_{\rm r}\ge W_{\rm r, min})} \, ,
\end{align}

\noindent where $\mathcal{N}_{\rm abs}$ is the actual number of absorption lines with rest-frame equivalent width, $W_r$,\footnote{The rest-frame equivalent width is related to the observed equivalent with by $W_{\rm r} = W_{\rm obs}/(1+z)$. } larger than a minimum threshold, $W_{\rm r, min}$, found within a given ``redshift path'', $\Delta z$, defined as the redshift range where absorption lines of $W_{\rm r}>W_{\rm r, min}$ could have been observed if these were present. The redshift path corresponds to the (1-dimensional) searching ``volume'', and it depends on $S/N$ and/or wavelength coverage of the quasar spectra; in contrast to surveys that rely on emission (e.g. galaxy surveys), the selection function for detecting absorbers within a quasar sightline is fairly uniform with redshift. $\mathcal{N}_{\rm abs}$ and $\Delta z$ can be constructed by summing up the contributions from the different quasar sightlines in a given survey, and they can be split into specific redshift ranges and/or intrinsic properties of the absorbers (e.g. column density).\sk

If we model absorbers as discrete ``clouds''\footnote{But note these do not need to have a spherical geometry. Indeed, at $1+\delta\lesssim 10$ we expect these to be distributed in sheet-like or filamentary structures following the dark matter `cosmic web', rather than halos.} that produce the individual absorption lines we see in the spectra of quasars, we can relate the \dndztt\ to their volumetric number density, $n_{\rm abs}(z)$, and (proper) cross-section, $\sigma_{\rm abs}(z)$, as
\begin{align}
\label{eq:dndz_2}
  \frac{{\rm d}\mathcal{N_{\rm abs}}}{{\rm d}z} = \frac{c}{H_0}n_{\rm abs}(z) \sigma_{\rm abs}(z) \frac{{\rm d}X}{{\rm d}z} \, ,
\end{align}

\noindent  where ${\rm d}X/{\rm d}z$ is the differential ``absorption distance'' defined below.\footnote{We note that Equation~\ref{eq:dndz_2} assumes unity covering fraction, but this could be easily relaxed by adding a $f_{\rm c}$ term representing it.} If we observe a redshift evolution in \dndztt, it could be intrinsic to the absorbers themselves and/or simply due to the expansion of the Universe. In order to account for the latter, ${\rm d}X/{\rm d}z$ relates the proper absorption distance to the redshift path due to the Universe's expansion \citep{bahcall1969},
\begin{align}
    \frac{{\rm d}X}{{\rm d}z} \equiv \frac{(1+z)^2}{\sqrt{\Omega_{\rm m}(1+z)^3 + \Omega_{\Lambda}}} \, ,
\end{align}
\noindent for a flat \lcdm\ Universe at late times, such that 
\begin{align}
\label{eq:dndz_3}
  \frac{{\rm d}\mathcal{N_{\rm abs}}}{{\rm d}z} = \mathcal{N_{\rm abs}}(z) \frac{{\rm d}X}{{\rm d}z} \, ,
\end{align}

\noindent where $\mathcal{N_{\rm abs}}(z) = \frac{c}{H_0}n_{\rm abs}(z) \sigma_{\rm abs}(z)$. This term can be parametrized as a power-law of the form,
\begin{align}
\label{eq:dndz_4}
    \mathcal{N_{\rm abs}}(z) = \mathcal{N_{\rm abs,0}}(1+z)^{\epsilon} \, ,
\end{align}
\noindent such that Equation~\ref{eq:dndz_2} can be written as,
\begin{align}
     \frac{{\rm d}\mathcal{N_{\rm abs}}}{{\rm d}z} =\mathcal{N_{\rm abs,0}} \frac{(1+z)^{2+\epsilon}}{\sqrt{\Omega_{\rm m}(1+z)^3 + \Omega_{\Lambda}}} \, .
\end{align}

\noindent Modeling the observed \dndztt\ in this manner, one can establish the value of $\epsilon$. If $\epsilon \approx 0$ then the absorbers are not evolving significantly. Here we note that some studies parametrize \dndztt\ as\footnote{For early studies, this parametrization made sense as the \lcdm\ cosmology was not established and thus $\Omega_{\rm \Lambda}=0$.}
\begin{align}
     \frac{{\rm d}\mathcal{N_{\rm abs}}}{{\rm d}z} =\mathcal{N_{\rm abs,0}} (1+z)^{\gamma_{n}} \, ,
\end{align}

\noindent instead (e.g. \citealt{janknecht2006, kim2021}) but the main idea remains the same: by measuring $\gamma_{n}$ one can establish if absorbers are intrinsically evolving or not. These studies have revealed that at $z>2$, $\mathcal{N_{\rm abs}}(z)$ shows significant evolution, with $\gamma_n>1-2$, whereas at $z\lesssim 1$, \dndztt\ is consistent with following the expectation from cosmic expansion \citep{janknecht2006, kim2021}. An evolution in $\mathcal{N_{\rm abs}}(z)$ can be interpreted as an evolution of either the number density or the cross-section of the absorbers, as this technique is only sensible to the product of $n_{\rm abs}(z)\sigma_{\rm abs}(z)$.\sk

\begin{figure}[t]
\centering
\includegraphics[width=0.6\textwidth]{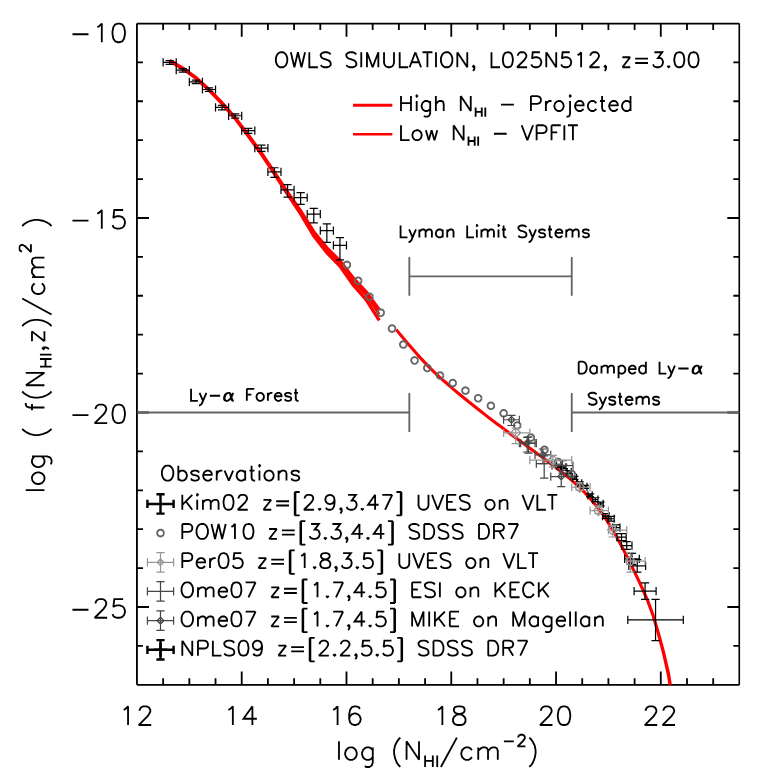}
\caption{Model and observational constraints for the \hi\ column density distribution at $2 \lesssim z \lesssim 5$. The model of \citet{altay2011} is shown by the red curve, based on a cosmological hydrodynamical simulation evaluated at $z=3$. Figure from \citet{altay2011}.}
\label{fig:altay}
\end{figure}

\paragraph{The \hi\ column density distribution}\sk

As seen before, not all \hi\ absorbers probe the same medium, as their properties will depend on density, temperature, ionization condition, etc. Empirically, we observe that low-\nhi\ absorbers are much more common than high-\nhi\ absorbers. Based on the previous analysis this could imply that low-\nhi\ absorbers are either more numerous and/or have larger cross-sections than their high-\nhi\ counterparts. Independently of the interpretation, the column density distribution of \hi\ is a key observable that has been widely used to shed light into the nature of absorbers, and their importance in cosmology and astrophysics.\sk

The \hi\ column density distribution is defined as
\begin{align}
    f(N_{\rm HI}, z) \equiv \frac{\partial^2 \mathcal{N_{\rm abs}}}{\partial N_{\rm HI} \partial X} 
    = \frac{\partial^2 \mathcal{N_{\rm abs}}}{\partial N_{\rm HI} \partial z} \frac{{\sqrt{\Omega_{\rm m}(1+z)^3 + \Omega_{\Lambda}}}}{(1+z)^2} \, ,
\end{align}

\noindent again, the $({\rm d}X/{\rm d}z)^{-1}$ term ensures that $f(N_{\rm HI}, z)$ is constant if $\mathcal{N_{\rm abs}}$ does not evolve with $z$. $f(N_{\rm HI}, z)$ can be inferred directly from quasar observations by establishing their \dndztt\ as a function of \nhi. Typically, $f(N_{\rm HI}, z)$ is parametrized as a power-law of the form,
\begin{align}
    f(N_{\rm HI}, z) = C N_{\rm HI}^{-\beta_{n}} (1+z)^{\gamma_n}\, ,
\end{align}
\noindent with $C$ a constant and $\beta_n>0$, although the explicit dependency on $z$ is sometimes not given when the measurements are performed over narrow redshift ranges \citep{tytler1982,janknecht2006,danforth2016}. \sk

The shape of $f(N_{\rm HI}, z)$ can be used to constrain $\Gamma_{\rm HI}$ (e.g. via Equation~\ref{eq:NHI_jeans2}) and is also important for estimating the total mass budget in the IGM traced by \hi\ absorbers (see below). Depending on the range of \nhi\ covered, and/or the statistical precision in the measurements, deviations from a single power-law have been reported; indeed, whether a single or a multiple power-law model (with different $-\beta_n$ slopes) is required to fit the $f(N_{\rm HI}, z)$ can be interpreted as evidence for single or multiple populations of \hi\ absorbers \citep{tytler1982,prochaska2010, kim2021}. \sk

To go beyond concatenations of different power-laws, it is convenient to rely on cosmological hydrodynamical simulations. Figure~\ref{fig:altay} shows a compilation of several observational works that estimate $f(N_{\rm HI}, z)$ at $z\gtrsim 3$ for different ranges of \nhi, together with a modeling based on a cosmological hydrodynamical simulation at $z=3$ presented by \citet{altay2011}. Their model includes ionization corrections and self-shielding \citep{zheng2002} that affect the absorbers at \nhi$\gtrsim 10^{17.2}$\,\cm~(LLSs), as well as the transition to a mostly neutral medium at higher \nhi$\gtrsim 10^{20.3}$\,\cm~(DLAs). At even higher \nhi$\gtrsim 10^{21.5}$\,\cm, they also model the effects of molecular hydrogen gas ($H_2$) and star formation. These results shed light into the origin of the \hi\ absorbers, where changes in the slope of the $f(N_{\rm HI}, z)$ can be attributed to changes in the physical conditions of the absorbing gas. The remarkable agreement between observations and theory (over $10$ orders of magnitude in \nhi!) indicates that our general understanding of the physical origin of \hi\ absorbers is robust.\sk

\paragraph{Baryon budget of the \hi\ absorbers}\sk

At a given redshift, the average mass density traced by \hi\ absorbers can be obtained by integrating the $f(N_{\rm HI}, z)$ over all \hi\ column densities and correcting for ionization. Parametrizing this gas density in units of the critical density of the Universe, $\rho_{\rm c}$, we have \citep{schaye2001} 
\begin{align}
    \Omega_{\rm g}(z) = \frac{8 \pi G m_{\rm H}}{3H_0c(1-Y)} \int \frac{n_{\rm H}}{n_{\rm HI}} N_{\rm HI} f(N_{\rm HI}, z)\, {\rm d}N_{\rm HI} \, .
\end{align}
\noindent where $Y$ is the baryonic mass fraction in helium ($Y=0.24$). Restricting this analysis to the diffuse IGM, and assuming the absorbers are in local hydrostatic equilibrium (Section~\ref{sec:equilibrium}), then combining Equations~\ref{eq:case_A},~\ref{eq:neutral_frac}~and~\ref{eq:NHI_jeans1} imply \citep{schaye2001} 
\begin{align}
    \Omega_{\rm g}(z) \approx 3.1 \times 10^{-9} 
    \left( \frac{\Gamma_{\rm HI}}{10^{-12}\ {\rm s}^{-1}} \right)^{1/3}
    \left(\frac{f_{\rm g}}{0.17} \right)^{1/2}
    \left(\frac{T}{10^4\,{\rm K}} \right)^{0.59}
    \int  N_{\rm HI}^{1/3} f(N_{\rm HI}, z)\, {\rm d}N_{\rm HI} \, .
\end{align}

\noindent here we have assumed a value of $H_0=70\, {\rm km\,s}^{-1}{\rm \,Mpc}^{-1}$, and that the temperature does not vary strongly as a function of \nhi. Evaluating $f(N_{\rm HI}, z)$ and integrating over typical \lya~forest lines at $z\gtrsim 3$ provide $\Omega_{\rm g} \sim \Omega_{\rm b}$, implying that the vast majority of baryons are in the diffuse IGM at these high redshifts \citep{fukugita1998, schaye2001}. However, the same calculation at $z\lesssim 0.5$ only accounts for $\Omega_{\rm g} \sim 0.3\Omega_{\rm b}$ \citep{shull2012}. Intriguingly, condensed matter in the CGM or within galaxies (cold gas, dust, stars, etc.) and hot gas in galaxy clusters (e.g. traced by $X$-ray emission) at $z\sim 0$ only account for $\sim 0.1\Omega_{\rm b}$ \citep{bregman2007, shull2012}. This $\sim 50$\,\% deficit of baryons in the low-$z$ Universe is known as the `missing baryon problem' \citep{fukugita1998}. According to hydrodynamical simulations, these missing baryons are in the WHIM (\citealt{cen1999, dave2001,cen2006,shull2012}; see Figure~\ref{fig:phase_diagram}), which have eluded a firm observational confirmation from our traditional techniques (quasar absorption lines and/or $X$-ray emission) due to their lack of sensitivity. However, complementary observational techniques are now allowing us to directly detect all these baryons (see Section~\ref{sec:other_techs}).\sk

\begin{figure}[t]
\centering
\includegraphics[width=0.9\textwidth]{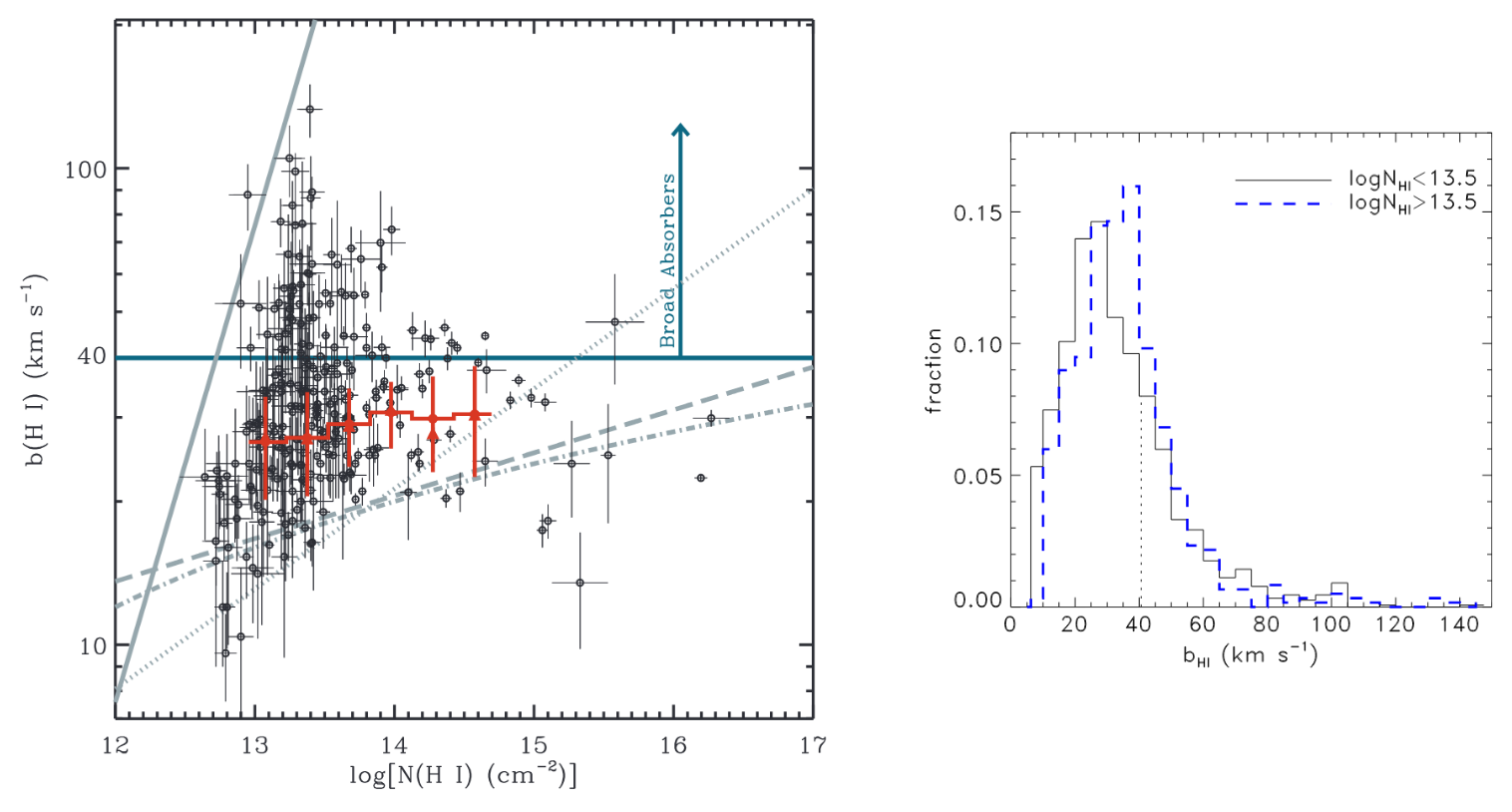}
\caption{Left: A Doppler parameter versus column density diagram for \hi\ absorbers at $z \lesssim 0.4$ from \citet{lehner2007}. The solid line marks the observational limit of their survey, such that lines at the left of this line are missed. The dotted grey line shows the predicted minimum $b_{\rm HI}$ according to hydrodynamical simulations. The dashed and dotted-dashed lines show the observed lower envelope for a sample of high redshift \hi\ absorbers (see references in \citealt{lehner2007}). The horizontal solid line represent the typical limit between broad and narrow \hi\ absorbers at $b_{\rm HI}=40$\,\kms\ (corresponding to $T=10^5$\,K assuming no turbulence). Finally, the red points show the mean (circles) and median (triangles) $b_{\rm HI}$ values for six bins in \nhi\ restricting to just absorbers with $b<40$\,\kms. Figure from \citet{lehner2007}. Right: \hi~Doppler parameter distribution for \nhi$<10^{13.5}$\,\cm\ (solid black) and \nhi$>10^{13.5}$\,\cm\ (dashed blue) absorbers at $z\lesssim 0.5$. Figure from \citet{danforth2016}.}
\label{fig:lehner}
\end{figure}

\paragraph{The \hi\ Doppler parameter distribution}\sk

Another important observable from \hi\ absorbers is the distribution of Doppler parameters, $b_{\rm HI}$. Doppler parameters encode thermal and/or non-thermal broadening (e.g. turbulence), characterized as (see Section~\ref{sec:qal_model_deeper})
\begin{align}
    b_{\rm HI}=\sqrt{\frac{2kT}{m_{\rm H}} + b^{2}_{\rm turb.}} \, .
\end{align}
\noindent Therefore, we can use $b_{\rm HI}$ to constrain the temperature of the IGM, given some assumptions or estimations for $b_{\rm turb.}$. For instance, one can treat the inferred $T$ from $b_{\rm HI}$ as a lower limit (i.e. there is no knowledge of $b_{\rm turb.}$) and fit for a density-temperature relation (see Section~\ref{sec:thermal}) in the lower envelope of a $b_{\rm HI}$-\nhi\ diagram (\citealt{schaye2000, kim2001, lehner2007}; see Figure~\ref{fig:lehner}, left panel). Alternatively, one can use metal ion absorbers that are aligned (kinematically) with the \hi\ and disentangle $b_{\rm turb.}$, by assuming that the both the \hi\ and the metal ion absorption are probing the same gas phase (e.g. \citealt{savage2014}; although see \citealt{marra2024}). Even for non-resolved absorption lines (especially relevant for the crowded \lya\ forest at high-$z$), one can still use the shape of the line-profile at the line centers to constrain $T$ \citep{lidz2010}. Empirically, the $b_{\rm HI}$ distribution is well characterized by a log-normal function, with a peak at around $b_{\rm HI}^{\rm peak} \sim 20-30$\,\kms\ and an extended tail towards larger values, $b_{\rm HI}^{\rm tail} \sim 100$\,\kms\ (\citealt{lehner2007, danforth2016}; see Figure~\ref{fig:lehner}, right panel). These kind of measurements indicate that the temperature of the diffuse IGM at mean density is of order $T\approx 0.5-2 \times 10^4$\,K, depending on redshift (e.g. \citealt{schaye2000}; see Figure~\ref{fig:temperature}).\sk

Assuming no turbulent broadening, then $T>10^5$\,K would correspond to $b_{\rm HI}> 40$\,\kms; therefore, the tail at $b_{\rm HI}\gtrsim 40$\,\kms\ could be partly due to the WHIM \citep{richter2006, danforth2010}. However, empirically establishing the temperature for these broad \hi\ absorbers (BLAs) has turned out to be very difficult. Apart from the uncertainties associated with estimating $b_{\rm turb.}$, these absorbers require higher $S/N$ and higher resolution given their intrinsically smaller $\tau_0$ (for a fixed \nhi; see Equation~\ref{eq:tau0}) and are more likely to blend with nearby systems \citep{richter2006}. Still, there has been observational evidence of a possible excess of BLAs in inter-cluster filaments, which could be an indirect signature of the WHIM (\citealt{tejos2016}).\sk

\paragraph{The connection between \hi~\lya\ absorbers and galaxies}\sk

According to the \lcdm\ paradigm, galaxies and gas in the IGM are both distributed following the same cosmic web shaped by dark matter and dark energy. Yet, testing this prediction has been challenging, because this requires having sufficient data on galaxies and IGM absorbers in the same volumes and over Mpc scales. At $z\lesssim 0.5$, where we have a relatively good mapping of galaxies, we require (restrictive) space-based UV spectroscopy to access \hi. In contrast, at $z\gtrsim 2$, we have a good mapping of \hi\ via optical spectroscopy, but we lack sufficient mapping of galaxies. Early studies at $z\lesssim 0.5$ established a spatial correlation between \hi~and galaxies, leading to two main interpretations: i) \hi\ absorbers are part of extended galaxy halos \citep[e.g.][]{lanzetta1995}, and ii) \hi\ absorbers are correlated with galaxies mostly because they both trace the same underlying matter distribution but are not necessarily the same physical systems \citep[e.g.][]{morris1993}. Later studies confirmed that the relative importance of these two interpretations depends on \nhi: at \nhi$\gtrsim 10^{15}$\cm, the absorbers are mostly related to the CGM (i.e. galaxy halos), whereas at \nhi$\lesssim 10^{15}$\cm, the absorbers are mostly tracing the diffuse IGM in the large-scale structure \citep{chen2005, prochaska2011, tejos2014, burchett2020}. Similar conclusions can be reached from observations at $z\gtrsim 2$ \citep{adelberger2003, adelberger2005, rudie2012}. Interestingly, observations have also established the existence of a population of \hi\ absorbers that reside within galaxy voids \citep{penton2002, tejos2012}, which are not correlated with galaxies at all \citep{tejos2014}.

\paragraph{Metals and enrichment of the IGM}\sk

Although in this Chapter we have focused on mostly on \hi\ absorbers, some of the methodologies presented can also be applied to other ions, including metals. The metal enrichment of the IGM is an important observable for constraining galaxy evolution models, as metals must originate within galaxies \citep{theuns2002, scannapieco2002, oppenheimer2006}. Current observations of resolved metal absorption lines are mostly tracing CGM or galaxy environments \citep{ellison2000,tripp2008, dodorico2016, peroux2020} but some intervening high-ionization ions like \ovi, \ovii~and~\nv\ may be tracing some (high-density) regions of the WHIM \citep{tumlinson2017, nicastro2018, artale2022}. Similar conclusions can be reached from studies from unresolved lines \citep{aracil2004}. We refer the reader to Chapters~\chcgm, \chgalwinds\ and \chfirstgals\ for more discussion on metal ions.\sk

\paragraph{Global pixel-by-pixel statistics}\sk

When individual absorption lines are not resolved, it is still possible to obtain relevant information from quasar spectroscopy. One such example is the Gunn-Peterson Effect described earlier. Similarly, under the assumption that the opacity in a given pixel is dominated by \hi\ absorbers in the highly ionized diffuse IGM (\lya-forest) in PIE (Equation~\ref{eq:neutral_frac}) at a single density and temperature (Equation~\ref{eq:tau_nu_constant}) following the temperature-density relation (Equation~\ref{eq:Tdensity}), we have
\begin{align}
    \tau_{{\rm Ly}\alpha} & \propto n_{\rm HI} \propto n_{\rm H}^2 \alpha_A(T) \Gamma_{\rm HI}^{-1}  \propto n_{\rm H}^2  T^{-0.71} \Gamma_{\rm HI}^{-1} \propto n_{\rm H}^{1.58}\Gamma_{\rm HI}^{-1} \,\\
    \tau_{{\rm Ly}\alpha} & \propto (1+\delta)^{1.58} \, ,
\end{align}

\noindent meaning that fluctuations in $\tau_{\nu}$ can be used to estimate fluctuations in the underlying density field. This is known as the `fluctuating Gunn-Peterson approximation' and is widely used to test models of structure formation and infer physical parameters of the IGM (see Chapter~\chcosmolya). Other important measurements include the power-spectrum and the mean transmission flux distribution of the \lya-forest, which can also be used to constrain the the nature of dark matter and test our cosmological paradigm \citep{croft1998,zaldarriaga2001, viel2004, mcdonald2005, becker2013}. We refer the reader to Chapters~\chlinear, \chpowerspectrum, \chbao\ and~\chcosmolya\ for more details on cosmological applications.\sk


\section{Concluding remarks and future prospects}
\label{sec:final}


The IGM remains indispensable for shaping our understanding of cosmology and galaxy evolution. As the reservoir of most of the baryonic matter in the Universe, it is fundamental to the processes that govern large-scale structure formation and the baryon cycle of galaxies. Although the physics of the IGM is relatively well understood, there are still key questions that are important to address in the nearby future. Some of these include:\sk

\begin{itemize}
        \item Where are the `missing baryons'? 
        \item Is there really a WHIM, as hydrodynamical simulations predict? 
        \item If the WHIM exists, what are its typical densities, temperatures, and environments?
        \item How important are galaxy feedback processes in injecting energy to the WHIM and the rest of the IGM?
        \item What is the metallicity of the IGM (beyond galaxy halos) as a function of cosmic environment and redshift?
        \item What are the sources of the UVB that ionized the IGM during and after EoR?
        \item What are the sources of the UVB that ionzed \heii\ at $z\sim 3$?
        \item What is the shape of the baryonic cosmic web in relation to that of dark matter (beyond 2-point statistics)? Are baryons distributed in a one-to-one correspondence with dark matter? Does this still hold on scales closer to that of galaxy halos? 
        \item What is the relation of the underdense IGM and galaxies in the cosmic web (e.g. galaxy voids)?
        \item What are the strengths and relative importance of magnetic fields? \sk
        \end{itemize}

Quasar spectroscopy has been fundamental for establishing our current knowledge of the IGM, but this technique alone may not be enough to answer all these pressing questions. New discoveries and new technologies have enabled the development of complementary observational techniques that can help address these and future questions regarding the IGM.

\subsection{Other complementary observational techniques}\sk
\label{sec:other_techs}

For advancing our understanding of the IGM, we must rely on complementary techniques to that of quasar spectroscopy. In the following, we briefly mention some of these:\sk 

\begin{itemize}

     \item{\bf Probing the IGM with fast radio bursts (FRBs):} FRBs are extragalactic millisecond radio pulses of yet unknown origin (\citealt{lorimer2007, cordes2019}; see Chapter~\chfrbs). The dispersion measure of FRBs is DM$=\int n_e \, {\rm d}s$, and thus these pulses can be used to directly trace ionized baryons in the IGM, independent of temperature. This technique requires isolating the intergalactic DM contribution out of the total observed DM. A recent application of this concept has confirmed that all the expected baryons are indeed in a highly ionized medium, thus conclusively solving the `missing baryon' problem (\citealt{macquart2020}; see also Chapter~\chcosmofrb). However, the exact location of these baryons in relation to the cosmic web and/or galaxy halos (e.g. IGM versus CGM) is still unknown; addressing this issue requires a sufficiently complete mapping of large-scale structures towards FRB sightlines \citep{khrykin2024}. Moreover, FRBs can also constrain the magnetic fields via their rotation measures, and also have the potential to constrain \heii\ reionization if observed at $z\gtrsim 3$ (see Chapter~\chcosmofrb).\sk
    
    \item {\bf Detecting the WHIM with the thermal Sunyaev-Zeldovich (tSZ) effect:} The tSZ effect is produced by Compton scattering of CMB photons that interact with electrons in a hot medium \citep{sunyaev1972}. This technique provides a signal that is $\propto \int n_e T \, {\rm d}s$ of the intervening gas, and therefore could be used to detect and characterize the WHIM, assuming there is independent knowledge of $n_e$. Recent studies have detected a positive signal from the stacking of the tSZ maps over large samples of massive galaxy pairs that could be intergalactic \citep{degraaff2019}. These authors combined their tSZ signal with weak lensing mass estimates to infer that their stacked signal comes from overdensities of $1+\delta\approx 5$ and a temperature close to $T\approx 3\times 10^{6}$\,K, i.e. comparable to those of a WHIM. Future higher resolution tSZ maps (e.g. \citealt{ade2019}), in combination with weak-lensing mass estimates, can thus provide further evidence for the WHIM.\sk
    
    \item {\bf Probing the IGM with $X$-ray emission:} In the low redshift Universe, one of the best prospects for detecting the IGM in emission is via $X$-ray imaging of the WHIM. Bremsstrahlung (free-free emission) in a hot medium produce $X$-ray photons, as commonly observed in the ICM of galaxy clusters (see Chapter~\chclusters). This emission is $\propto n_e n_{\rm HII} \sqrt{T}$, i.e. $\propto n_{\rm H}^2\sqrt{T}$ in a highly ionized medium. The temperature can be determined via the frequency cut-off of the emission, thus $X$-ray imaging and spectroscopy can be used to constrain both the density and temperature of the gas. With current sensitivities, some detections of IGM material in inter-cluster filaments have already been reported with inferred overdensities of $1+\delta \approx 150$ and temperatures $T\sim10^7$\,K \citep{werner2008}. These are at the limit between WHIM and `Hot Gas' phases, and is still unclear whether the CGM of massive galaxies or groups are affecting the results. Current and next generation of $X$-ray missions (including the recent eROSITA survey; \citealt{merloni2024, zhang2024}) will allow us to push the limits towards densities and temperatures more comparable with those of the WHIM. \sk

    \item {\bf Probing the IGM with \lya-emission:} At higher redshifts, probably the best strategy for observing the IGM in emission is through \lya~emission itself. Although this is beyond current capabilities for typical column densities of the \lya-forest absorbers, it is expected that the fluorescent \lya~emission produced by the outer layers of optically thick self-shielded clouds (e.g. LLS) may be within reach. Indeed, \citet{cantalupo2014} reported such an emission at intergalactic distances aided by the extra radiation field of a nearby quasar at $z\sim 2.3$. Whether the signal is truly intergalactic or not is still unclear. For instance, \citet{gallego2018} stacked on pairs of \lya-emitters (galaxies) at $3<z<4$ to search for this signal and could only detect a signal from the CGM of these galaxies rather than being intergalactic \citep{wisotzki2018}. Still, future $30$-m will have the sensitivities to attempt this experiment more efficiently.\sk

    \item {\bf Probing the IGM with multiple sightlines:} It is possible to extend the traditional quasar absorption line spectroscopy technique (limited to probing gas only in the line-of-sight direction) to study correlations in the transverse direction by using multiple quasars pairs, either binaries or strongly lensed systems. This has been already fundamental to measure the coherence lengths and establish the three-dimensional distribution and geometries of \hi\ (and metal) absorbers \citep{dinshaw1994, smette1995,fang1996, rauch2005,rorai2013}. Similarly, gravitational-arcs as background sources can provide multiple and contiguous coverage over several $100$\,kpc$^2$ for intervening material \citep{lopez2018}. Moreover, the use of background galaxies as background sources allow us to probe the IGM with a much finer sampling that what quasars pairs can afford; this concept has been recently applied to provide the first \lya-forest tomography of the cosmic web at $z>2$ \citep{lee2018}. The next generation of $30$-m optical telescopes will be able to routinely achieve this kind of mapping \citep{japelj2019}.\sk

    \item {\bf Probing the IGM with gamma-ray bursts (GRBs):} GRBs are bright short-lived bursts of gamma-ray radiation from highly energetic cosmic explosions, often associated with the collapse of massive stars or the merger of compact objects like neutron stars (see Chapter~\chgrbs). GRBs optical afterglows can be used to probe the IGM in a similar fashion as quasars, i.e. via spectroscopic analysis of their intervening absorption lines. The novelty here is that these afterglows are transient, and thus they eventually fade away. This allows for more sensitive search of intervening galaxies that could be associated to \hi\ absorbers compared to those around quasars \citep{chen2005}. Moreover, the energetics of GRBs imply that we could detect them at $z \gtrsim 8$ \citep{salvaterra2009}, thus these could also help us to constrain the physics of the EoR.\sk
    
\end{itemize}

Achieving a comprehensive understanding of the IGM requires further observational data and refined theoretical models to capture its complex behavior across cosmic time and environments. By continuously studying and analyzing  the light from various distant objects, we can refine our understanding of the IGM and its role in the cosmic landscape.\sk

\begin{ack}[Acknowledgments]
 
We thank Joe N. Burchett, Sebasti\'an L\'opez, J. Xavier Prochaska, and Jessica Werk for their comments and suggestions on the manuscript. We are also grateful to Karen Mart\'inez--Acosta for providing the figures with the quasar spectra. Additionally, we thank Gabriel Altay, Renyue Cen, Charles Danforth, Daniela Gal\'arraga--Espinosa, Cameron Hummels, Vikram Khaire, Nicolas Lehner, Davide Martizzi, R\"udiger Pakmor, and Phoebe Upton--Sanderbeck for granting permission to use their figures as part of this Chapter.
\end{ack}

\seealso{\citet{rauch1998, meiksin2009, draine2011, mcquinn2016,tumlinson2017,peroux2020, faucher2023}.}

\bibliographystyle{Harvard}
\bibliography{els-igm}

\end{document}